\documentclass{aa}  
\usepackage{graphicx}
\usepackage[varg]{txfonts}
\usepackage{natbib}
\usepackage{longtable}
\usepackage{lscape} 
\usepackage{graphicx}
\usepackage{url}
\usepackage{amssymb}
\usepackage{longtable}

\begin{document}

   \title{The Seven Sisters DANCe\thanks{Based on service observations made with the William Herschel Telescope operated on the island of La Palma by the Isaac Newton Group in the Spanish Observatorio del Roque de los Muchachos of the Instituto de Astrof\'\i sica de Canarias.}}

  \subtitle{I. Empirical isochrones, luminosity, and mass functions of the Pleiades cluster}

   \author{H. Bouy \inst{1}
          \and E. Bertin\inst{2}
          \and L.M. Sarro\inst{3}
          \and D. Barrado\inst{1}
          \and E. Moraux\inst{4}
          \and J. Bouvier\inst{4}
          \and J.-C. Cuillandre\inst{5}
          \and A. Berihuete\inst{6}
          \and J. Olivares\inst{3,4}
          \and Y. Beletsky\inst{7}
}

   \institute{Centro de Astrobiolog\'\i a, Dpto de Astrof\'\i sica, INTA-CSIC, PO BOX 78, E-28691, ESAC Campus, Villanueva de la Ca\~nada, Madrid, Spain\\
         \email{hbouy@cab.inta-csic.es}
         \and 
         Institut d'Astrophysique de Paris, CNRS UMR 7095 and UPMC, 98bis bd Arago, F-75014 Paris, France 
         \and
         Depto. de Inteligencia Artificial , UNED, Juan del Rosal, 16, 28040 Madrid, Spain 
         \and
         UJF-Grenoble 1/CNRS-INSU, Institut de Plan\'etologie et d'Astrophysique de Grenoble (IPAG), UMR 5274, Grenoble, F-38041, France
         \and
         Canada-France-Hawaii Telescope Corporation, 65-1238 Mamalahoa Highway, Kamuela, HI96743, USA 
         \and
         Dept. Statistics and Operations Research, University of C\'adiz, Campus Universitario R\'\i o San Pedro s/n, 11510 Puerto Real, C\'adiz, Spain
         \and
         Las Campanas Observatory, Carnegie Institution of Washington, Colina el Pino, 601 Casilla, La Serena, Chile
            }

   \date{Received ; accepted}

 
  \abstract
{The DANCe survey provides photometric and astrometric (position and proper motion) measurements for approximately 2 million unique sources in a region encompassing $\sim$80~deg$^{2}$ centered on the Pleiades cluster.}
   {We aim at deriving a complete census of the Pleiades and measure the mass and luminosity functions of the cluster.}
   {Using the probabilistic selection method described in \citet{2014A&A...563A..45S}, we identified high probability members in the DANCe ($i\ge$14~mag) and Tycho-2 ($V\lesssim$12~mag) catalogues and studied the properties of the cluster over the corresponding luminosity range.}
   {We find a total of 2\,109 high-probability members, of which 812 are new, making it the most extensive and complete census of the cluster to date. The luminosity and mass functions of the cluster are computed from the most massive members down to $\sim$0.025~M$_{\sun}$. The size, sensitivity, and quality of the sample result in the most precise luminosity and mass functions observed to date for a cluster. }
   {Our census supersedes previous studies of the Pleiades cluster populations, in terms of both sensitivity and accuracy. }


   \maketitle
%

\section{Introduction}
As  one of the closest and youngest open clusters, the Pleiades has been extensively studied  over the past century. 
From stellar activity to multiplicity and lithium abundance, many topics have been addressed using this fascinating cluster as a cornerstone. One very important subject is the shape of the origin of the initial mass function, a key tool in understanding the formation and evolution of the stellar populations in different environments. Its computation requires an accurate description of the stellar (and substellar) census.

As a matter of fact, the study of the Pleiades as a group started much earlier. Both Claudius Ptolemy and Tycho Brahe identified a few dozen stars. Even Galileo Galilei did not do much better when he made the first telescopic observation of the cluster. Coming back to the 20th century, the stellar census expanded rapidly due to the increase in sky coverage and the depth. To name a few that are important because the naming convention and prefixes: 
\citet{Trumpler1921.1} and  \citet{Trumpler1921.2} for Tr; \citet{Hertzsprung1947.1} for "HII"; \citet{Artyukhina1969.1} for "AK";
\citet{Haro1982.1} for "HCG"; 
\citet{vanLeeuwen1986.1} for "Pels"; 
\citet{Stauffer1991.1} for "SK";
\citet{Hambly1993.1} for HHJ; 
\citet{Pinfield2000.1} for "BPL"; 
and \citet{Deacon2004.1} for "DH". %
The Pleiades has also been a privileged target for the search and study of substellar objects. The first confirmed brown dwarfs were in fact discovered in the Pleiades \citep{1995Natur.377..129R, 1996ApJ...458..600B}, extending the mass function beyond the hydrogen-burning limit. Several studies subsequently identified a growing number of substellar Pleiades members, providing hints that the mass function flattens or even increases at the very low mass end \citep[e.g.,][ and references therein]{2003A&A...400..891M, 2012MNRAS.422.1495L} and suggesting different (or additional) formation mechanisms.

\citet{2007ApJS..172..663S} have recently presented an exhaustive list of candidate and confirmed members
that come from more than a dozen independent surveys of the
Pleiades down to the limit of sensitivity of the 2MASS 6x Atlas and Catalogs. The total number of members and candidate members reported
in their work adds up to 1\,471 objects. More recently,
\citet{2012MNRAS.422.1495L} have used the UKIDSS GCS survey to search for
low mass members. The limited astrometric accuracy of the UKIDSS
survey and the inconsistent probability membership calculations used
in their study resulted in an incomplete and contaminated list of
members \citep{2014A&A...563A..45S}. 

In \citet{2014A&A...563A..45S}, we
presented a new probabilistic method based on a multivariate analysis
to derive reliable membership probabilities using the DANCe catalog of
\citet{2013A&A...554A.101B}. In this article, we present a new release of the DANCe-Pleiades catalog including new photometry and updated membership probabilities (Appendix~\ref{secondrelease}), as well as a new analysis of the Tycho-2 catalog using the same technique (decribed in Appendix~\ref{tycho_sel}), in order to also tackle the high mass end of the cluster. We used them to derive an improved list of high probability cluster members, described in section~\ref{sec:thelist}. Since the Pleiades has played a very important role in the history of brown dwarf searches, we assess different surveys again by looking for substellar members in Section~\ref{sec:BD}. In section~\ref{sec:lumteff}, we derive the luminosities and effective temperatures of the Pleiades members. In section~\ref{sec:pdmf}, we compute the luminosity and mass functions of the cluster and compare them to theoretical predictions. Finally, in section~\ref{sec:isochrones} we provide emprical isochrones of the Pleiades single-star locus in the $B_{\rm T}$,$V_{\rm T}$,$u$,$g$,$r$,$i$,$z$,$Y$,$J$,$H$, and $Ks$ filters. 

\section{Building the sample of members \label{sec:thelist}} 
Deriving a reliable luminosity function representative of the true cluster's population requires a good understanding of the selection biases. In an effort to cover the entire luminosity range, we combined candidate members from two complementary sources.
\begin{itemize}
\item[-] \emph{Middle and low luminosity end:} the DANCe survey is expected to be complete over the dynamic range between 14$\lesssim i\lesssim$23~mag (see Fig.~\ref{fig:sensitivity}). A new (second) release of the DANCe-Pleiades catalog \citep{2013A&A...554A.101B} is provided with the present article, and described in Appendix~\ref{secondrelease}. The main improvement consists in the addition of $gri$ photometry from the AAVSO Photometric All-Sky Survey DR7 (APASS), as well as our own new Y-band photometry for a small subsample of candidate members at the bright end, when no measurement was available. These two additions allow us to recover many members previously missed because of the lack of a good training set with complete photometric coverage at the bright end of the DANCe survey. In \citet{2014A&A...563A..45S}, we discussed various selection biases in the DANCe-Pleiades catalog relative to the choice of the variables used for the membership probability calculations.  As recommended in this article and for the rest of the present study, we use the membership probability defined using the RF2, and a selection threshold at probability greater than or equal to 75\%. A total of 2\,010 sources were selected. 
\item[-] \emph{High luminosity end:} the Tycho-2 survey provides astrometric and photometric measurements for 2.5 million sources brighter than $V\lesssim$12~mag. We complemented the Tycho-2 $B_{T},V_{T}$ photometry with APASS, 2MASS, and CMC-14 photometry, and applied the method described in \citet{2014A&A...563A..45S} to the corresponding catalog within the same area as the DANCe survey. The selection process -- and in particular the choice of variables and selection threshold -- are described in more detail in Appendix~\ref{tycho_sel}. We identified 207 candidate members, which overlap with the DANCe selection mentioned above.
\end{itemize}

The DANCe and Tycho-2 samples overlap over more than 3~mag and include 110 common sources. The overlapping region is believed to be incomplete because it corresponds to the limit of sensitivity of the Tycho-2 catalog and to the saturation limit (largely inhomogeneous over the field covered) of the DANCe survey. Even though the final catalog seems to cover the entire luminosity range, the domain between 12$\lesssim i\lesssim$14~mag must suffer from biases and incompleteness that are hard to quantify. Finally, a visual inspection showed that Alcyone and Electra had been missed by our selection. Both are indeed absent in the Tycho-2 catalog. For completeness, we added them to the final list of members, which include a total of 2\,109 high probability members.

\begin{figure}
   \centering
   \includegraphics[width=0.45\textwidth]{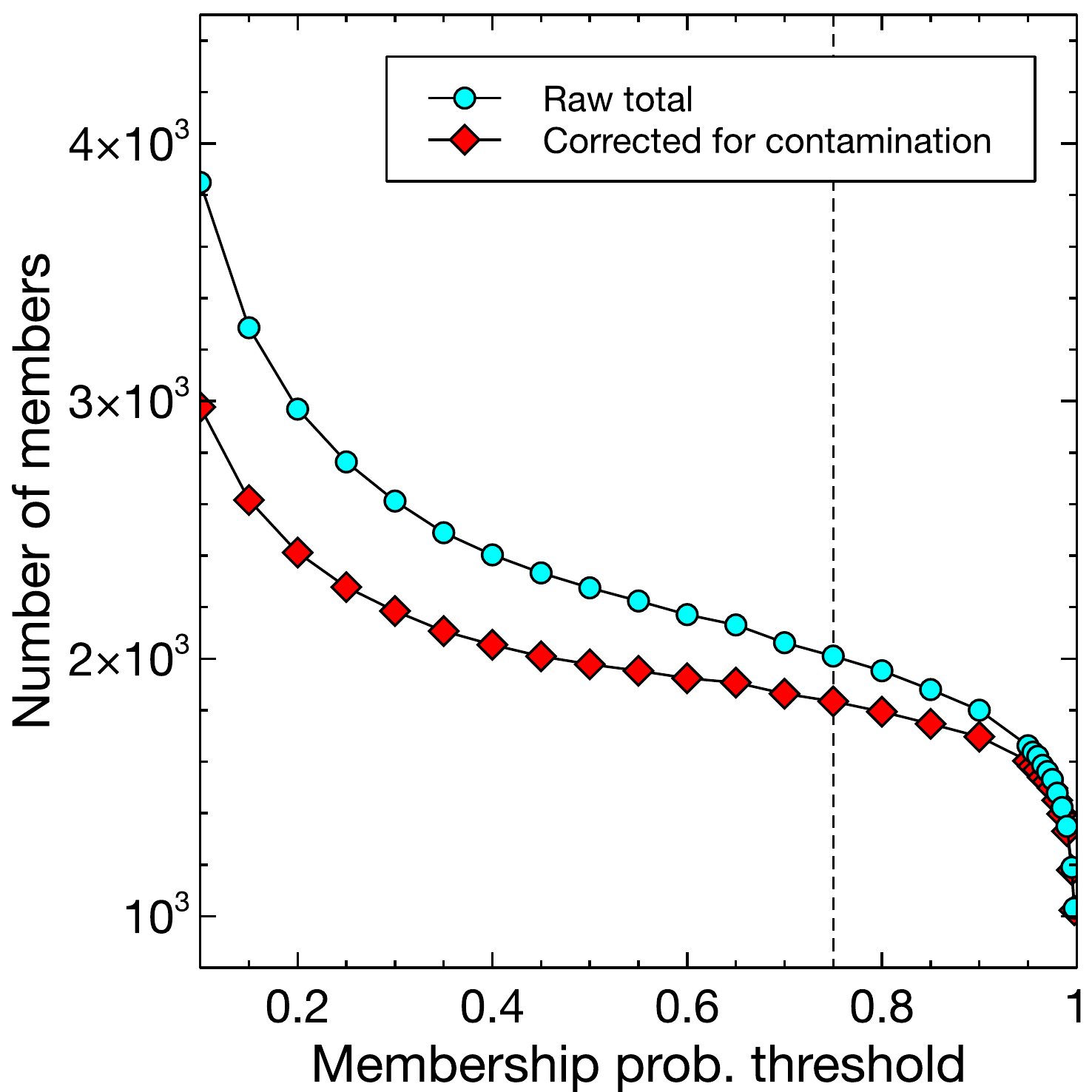}
      \caption{Sample size (not including the Tycho-2 analysis) as a function of the membership probability threshold. The raw (cyan) and corrected-for contamination (red) values are indicated. A dotted line indicates the 0.75 level chosen for the present study. }
         \label{fig:samplesize}
   \end{figure}

Figure~\ref{fig:samplesize} shows the sample size -- raw and corrected for contamination -- as a function of the membership probability threshold for the DANCe sample alone. The contamination rate was estimated using simulations as described in \citet{2014A&A...563A..45S}.  It shows that the total number of members after correction for contamination slowly increases to reach $\sim$2\,000 - 2\,100 members at a threshold of $\sim$0.2 - 0.3, and then rapidly increases at lower membership probabilities. The simulations probably largely underestimate the contamination level for the lowest membership probability values, explaining the sharp increase. But Fig.~\ref{fig:samplesize} nevertheless suggests that between 200 and 300 sources below the conservative 75\% threshold used in our analysis could be genuine members whose probability was lowered because of problematic photometric measurements or very large astrometric uncertainties. As such they deserve attention and follow-up observations to confirm their membership in the cluster. The same applies to the Tycho-2 sample, in which $\sim$10 - 20 members might have been missed by the 0.48 membership probability selection threshold (see Appendix~\ref{tycho_sel}). With this limitation in mind and to minimize the contamination level, for the rest of the present analysis we use the recommended 0.48 and 0.75 thresholds for the Tycho-2 and DANCe samples, respectively.

\section{Comparison with previous surveys}
\begin{figure}
   \centering
   \includegraphics[width=0.45\textwidth]{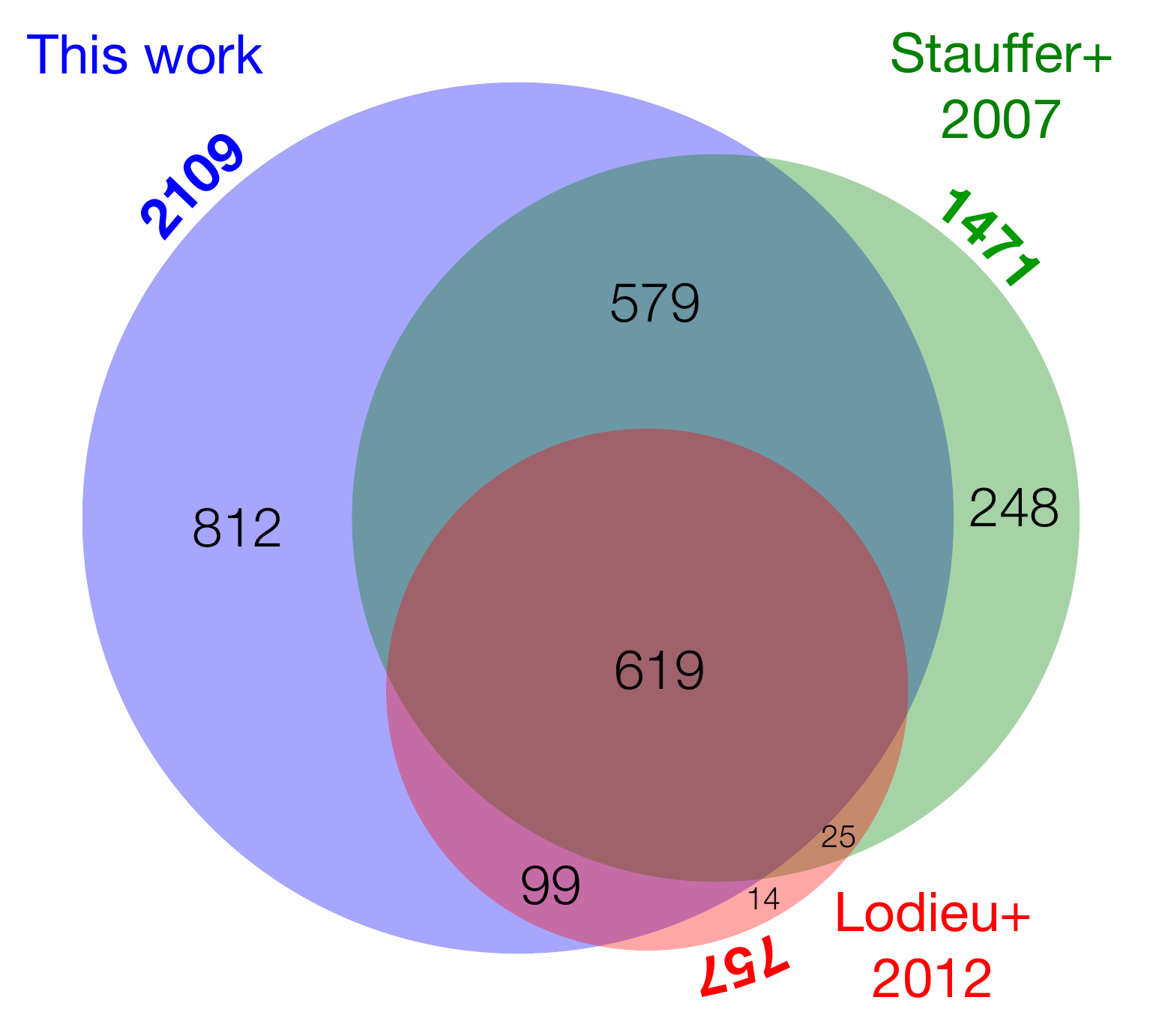}
      \caption{Venn diagram representing the sample of candidate members identified in the DANCe survey (blue), in \citet{2007ApJS..172..663S} (green), and in \citet{2012MNRAS.422.1495L} (red, probability greater than 50\% in their analysis). In each case the total number is indicated in bold face outside the circles.}
         \label{fig:venn}
   \end{figure}

\begin{figure}
   \centering
   \includegraphics[width=0.45\textwidth]{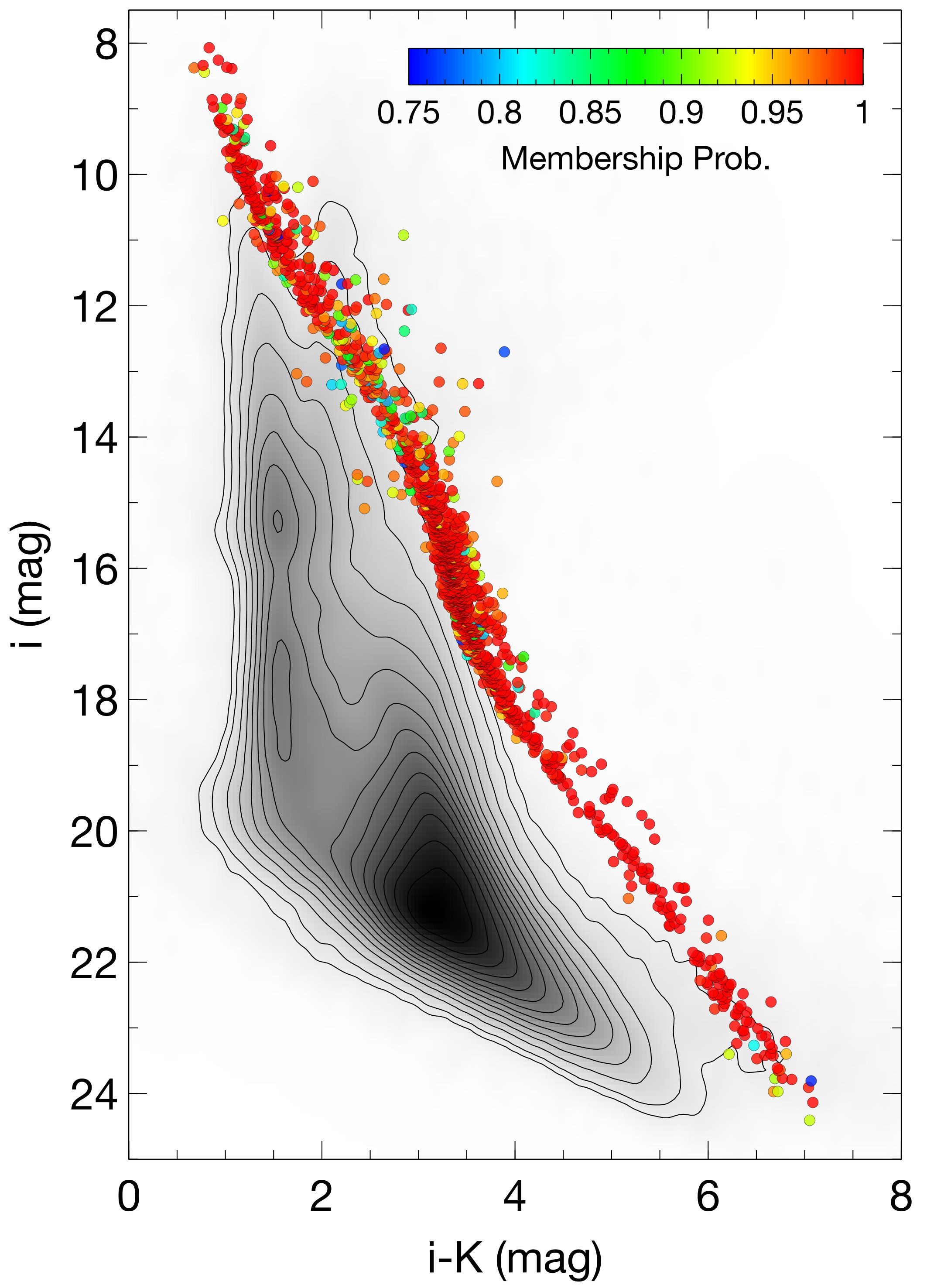}
      \caption{$i$ vs $i-K$ color-magnitude diagram of the DANCe survey. Sources with membership probabilities greater than or equal to 75\% are indicated as large colored dots. The color scale represents the membership probability. }
         \label{fig:i_iK}
   \end{figure}

We compare the result of our selection with the list of 1\,471 candidate members from \citet{2007ApJS..172..663S} and 757 candidate members from \citet{2012MNRAS.422.1495L}. These two studies cover a similar area, and their dynamic ranges largely overlap, resulting in a total of 644 candidates in common. The two lists combined add up to 1\,584 unique candidate members, of which 1\,297 have a counterpart within 2\arcsec\, in the DANCe catalog. Figure~\ref{fig:venn} shows a Venn diagram comparing the three samples of candidate members. A total of 812 candidate members identified in the present study are not included in \citet{2007ApJS..172..663S} or \citet{2012MNRAS.422.1495L}. Figure~\ref{fig:i_iK} shows a ($i$, $i-K$) color-magnitude diagram for the 1\,411 candidate members with both $i$ and $K$ photometry. 

\begin{figure}
   \centering
   \includegraphics[width=0.45\textwidth]{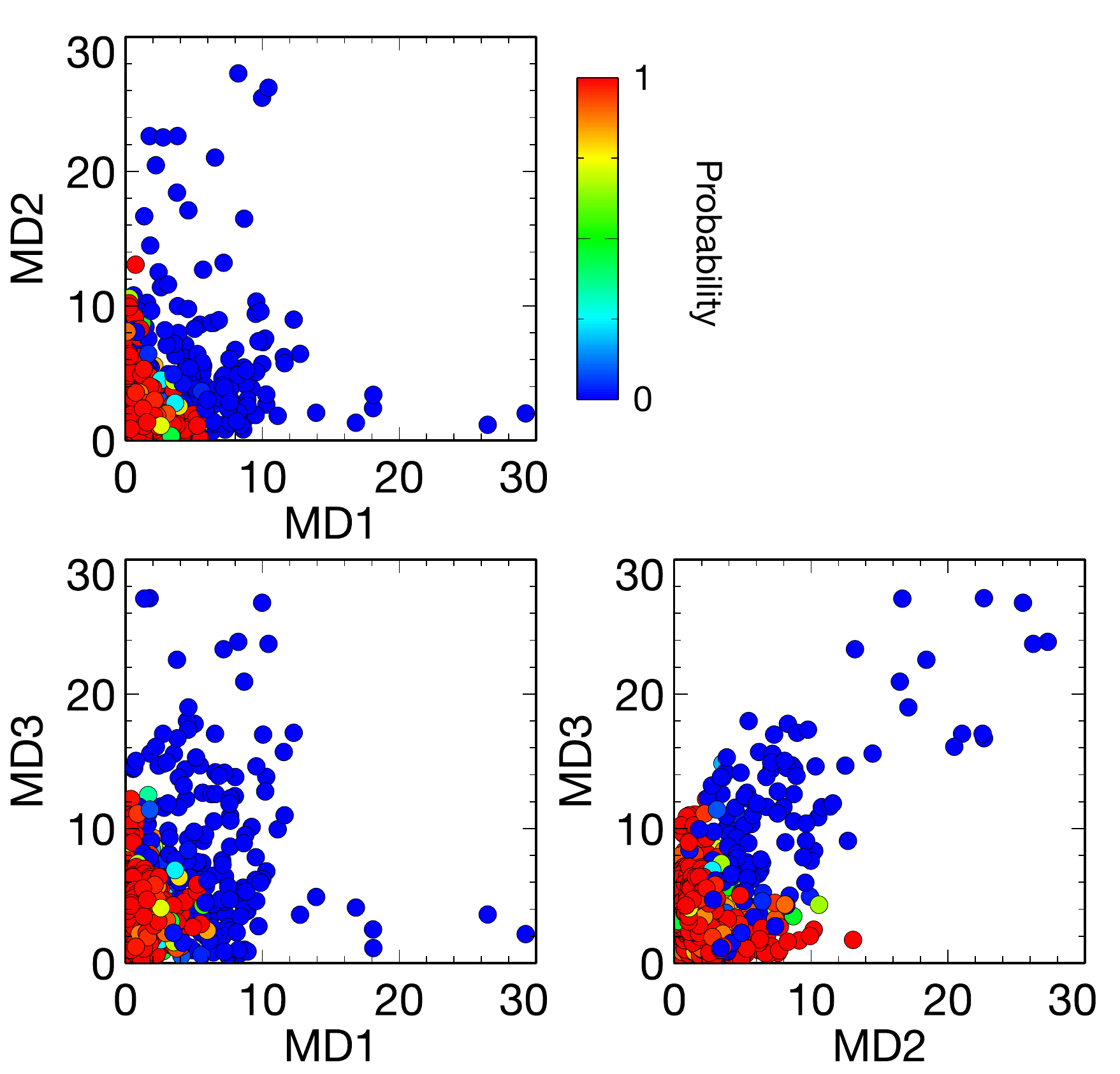}
      \caption{Distributions of Malahanobis distances for the sample of  \citet{2007ApJS..172..663S}. The color scale represents the membership probability as reported in the DANCe catalog.}
         \label{fig:mala2d_stauffer}
   \end{figure}

We investigated the candidates from  \citet{2007ApJS..172..663S} and \citet{2012MNRAS.422.1495L} with a counterpart in the DANCe survey but a membership probability below our 0.75 threshold. We explored the Malahanobis distances in the hepta-dimensional space of proper motions, luminosities, and colors. Figure~\ref{fig:mala2d_stauffer} presents the results. Four Malahanobis distances were explored: 
\begin{itemize}
\item MD1: minimum distance to any of the Gaussian components in the bidimensional proper motion space;
\item MD2: minimum distance to the principal curve (isochrone) of the single star locus in the penta-dimensional space made of colors and luminosities;
\item MD3: minimum distance to the principal curve (isochrone) of the equal-mass binary star locus in the penta-dimensional space made of colors and luminosities.
\end{itemize}
Each MD therefore tells whether a source has been rejected mostly because of its inconsistent proper motion (MD1) or photometry (MD2 and MD3) or both. Figure~\ref{fig:mala2d_stauffer} shows that in general and as one could expect, objects with low membership probabilities have large Malahanobis distances in several spaces, while high probability cluster members are generally located at short Malahanobis distances in all spaces. A source with an inconsistent proper motion (large MD1) is indeed most likely unrelated to the cluster so is likely to display colors and luminosities that are inconsistent with the cluster single and binary sequences (MD2 and MD3). These results suggest that in general the rejection is robust because it is consistent in several dimensions.

A number of members from \citet{2007ApJS..172..663S} have membership probabilities below our 0.75 threshold but above 0.1. As discussed above, these could well be genuine members as well. Additionally, a number of candidate members from \citet{2007ApJS..172..663S} fall in the domain of incompleteness between the Tycho-2 and DANCe analyses, are missed by the present study, and contribute to the difference illustrated in the Venn diagram.

\section{Brown dwarfs in the Pleiades\label{sec:BD}}
The Pleiades cluster has been one of the most popular brown dwarf hunting grounds. The study by \citet{2012MNRAS.422.1495L} is discussed above. In the following we compare the list of candidate substellar members of the Pleiades from \citet{2003A&A...400..891M},  \citet{2006A&A...458..805B}, \citet{2007MNRAS.378.1131C}, and \citet{2014A&A...568A..77Z} to the DANCe list of high probability members. Earlier studies exist but are superseded by the latest ones.

\subsection{\citet{2003A&A...400..891M}}
Based on proper motion ($I,Z$) photometric measurements, \citet{2003A&A...400..891M} have discovered 37 brown dwarf candidate members of the Pleiades. The entire dataset of images used by these authors was included in the DANCe survey, ensuring a complete spatial overlap with the corresponding study. Only 31 candidates have a counterpart in the DANCe catalog within 2\arcsec. A visual inspection of the CFH12K, MegaCam, and UKIDSS images shows that three of the six remaining (CFHT-PLIZ-5,14, 18) are not detected in the CFH12K, MegaCam, or UKIDSS images, so we interpret them as false detections. The other three (CFHT-PLIZ-23, 34, 37) are fast-moving sources that visibly move by several arcsec over the 13-year period and can be rejected as well. 

A total of 13 objects classified as members by \citet{2003A&A...400..891M} have membership probabilities below 10\% and are considered as contaminants. All of them have such very low membership probabilities mostly because of proper motions inconsistent with the cluster's mean motion. Table~\ref{tab:comparison} gives a summary of the 19 candidate members from \citet{2003A&A...400..891M}  rejected in our analysis.

\begin{longtab}
\scriptsize
\begin{longtable}{llllrrrlrl}
\caption{Comparison with previous Pleiades brown dwarf surveys \label{tab:comparison}}\\
\hline\hline
  \multicolumn{1}{c}{Moraux+2003} &
  \multicolumn{1}{c}{Bihain+2006} &
  \multicolumn{1}{c}{Casewell+2007} &
  \multicolumn{1}{c}{ZO+2024} &
  \multicolumn{1}{c}{RA} &
  \multicolumn{1}{c}{Dec} &
  \multicolumn{1}{c}{Sarro+2014} &
  \multicolumn{1}{c}{DANCe} &
  \multicolumn{1}{c}{proba} &
  \multicolumn{1}{l}{Comments} \\
\hline
  CFHT-PLIZ-1 &  &  &  & 57.774916 & 24.604747 & 5277684 & J03510596+2436170 & 0.1 & \\
  CFHT-PLIZ-2 &  & PLZJ-78 &  & 58.846125 & 24.818058 & 5296873 & J03552307+2449051 & 1.0 & \\
  CFHT-PLIZ-3 & BRB11 &  &  & 58.028 & 24.266877 & 5172848 & J03520671+2416006 & 1.0 & \\
  CFHT-PLIZ-4 &  &  &  & 55.4205 & 25.906388 & 5384327 & J03414090+2554242 & 1.0 & \\
  CFHT-PLIZ-5 &  &  &  & 58.408166 & 26.038797 &  &  &  & (1)\\
  CFHT-PLIZ-6 & BRB9 & PLZJ-9 &  & 58.479583 & 23.393447 & 5061235 & J03535512+2323361 & 1.0 & \\
  CFHT-PLIZ-7 &  &  &  & 57.050541 & 25.907888 & 5393326 & J03481212+2554283 & 0.046 & \\
  CFHT-PLIZ-8 &  &  &  & 55.75075 & 24.731147 & 5224304 & J03430017+2443525 & 1.0 & \\
  CFHT-PLIZ-9 &  &  &  & 56.146625 & 25.228427 & 5236418 & J03443516+2513429 & 1.0 & \\
  CFHT-PLIZ-10 &  &  &  & 57.937375 & 23.444297 & 5050329 & J03514494+2326395 & 1.0 & \\
  CFHT-PLIZ-12 & BRB14 &  &  & 57.856708 & 23.755877 & 5163704 & J03512557+2345214 & 1.0 & \\
  CFHT-PLIZ-13 &  &  &  & 58.768333 & 26.263699 & 4292417 & J03550436+2615503 & 1.0 & \\
  CFHT-PLIZ-14 &  &  &  & 58.384958 & 26.116999 &  &  &  & (1)\\
  CFHT-PLIZ-15 &  &  &  & 58.077666 & 24.074558 & 5170275 & J03521864+2404280 & 1.0 & \\
  CFHT-PLIZ-16 &  &  &  & 55.917875 & 24.50315 & 5227533 & J03434030+2430114 & 1.0 & \\
  CFHT-PLIZ-17 &  &  &  & 57.861208 & 23.502958 & 5051897 & J03512665+2330108 & 0.996 & \\
  CFHT-PLIZ-18 &  &  &  & 58.503999 & 24.914697 &  &  &  & (1)\\
  CFHT-PLIZ-19 &  &  &  & 59.068208 & 23.914288 & 5181049 & J03561638+2354514 & 0.999 & \\
  CFHT-PLIZ-20 & BRB15 & PLZJ-11 &  & 58.522375 & 23.566519 & 5169202 & J03540534+2333595 & 1.0 & \\
  CFHT-PLIZ-21 &  &  &  & 58.865249 & 25.827977 & 5406704 & J03552762+2549407 & 0.977 & \\
  CFHT-PLIZ-22 &  &  &  & 57.969624 & 26.875599 & 5535893 & J03515262+2652319 & 0.0 & \\
  CFHT-PLIZ-23 &  &  &  & 57.8895 & 24.170599 & 5161880 & J03513347+2410142 & 0.0 & \\
  CFHT-PLIZ-24 &  &  &  & 56.848666 & 26.016597 &  &  &  & (1)\\
  CFHT-PLIZ-26 &  &  &  & 56.202749 & 25.654866 & 5390235 & J03444868+2539175 & 0.0 & \\
  CFHT-PLIZ-28 & BRB18 &  &  & 58.558458 & 23.297608 & 5054020 & J03541406+2317521 & 1.0 & \\
  CFHT-PLIZ-29 &  &  &  & 57.438708 & 26.847188 & 5518534 & J03494524+2650503 & 0.0 & \\
  CFHT-PLIZ-30 &  &  &  & 57.941666 & 26.827058 & 4177219 & J03514595+2649382 & 0.0 & \\
  CFHT-PLIZ-31 &  & PLZJ-21 &  & 57.948541 & 24.66653 & 5285380 & J03514765+2439591 & 1.0 & \\
  CFHT-PLIZ-32 &  &  &  & 57.564458 & 26.580908 & 5517035 & J03501543+2634515 & 0.0 & \\
  CFHT-PLIZ-33 &  &  &  & 57.686166 & 26.702599 &  &  &  & (1)\\
  CFHT-PLIZ-34 &  &  &  & 58.510666 & 24.673908 & 5295295 & J03540255+2440259 & 0.0 & \\
  CFHT-PLIZ-35 & BRB20 &  &  & 58.163208 & 24.775008 & 5290268 & J03523915+2446296 & 1.0 & \\
  CFHT-PLIZ-36 &  &  &  & 58.65975 & 23.633508 & 5169338 & J03543837+2338011 & 0.0 & \\
  CFHT-PLIZ-37 &  &  &  & 58.914875 & 24.214477 &  &  &  & (1)\\
  CFHT-PLIZ-38 &  &  &  & 56.477875 & 26.504047 & 5512940 & J03455470+2630149 & 0.0 & \\
  CFHT-PLIZ-39 &  &  &  & 58.417916 & 26.271708 & 4181633 & J03534029+2616198 & 0.0 & \\
  CFHT-PLIZ-40 &  &  &  & 57.455416 & 26.565608 & 5518403 & J03494925+2633563 & 0.0 & \\
   & BRB1 &  &  & 57.964999 & 23.580611 & 5161183 & J03515155+2334492 & 1.0 & \\
   & BRB2 &  &  & 58.184583 & 23.904222 & 5171639 & J03524428+2354151 & 0.999 & \\
   & BRB3 &  &  & 58.032916 & 23.987388 &  &  &  & (1)\\
   & BRB4 & PLZJ-29 &  & 56.096666 & 25.645749 & 5385058 & J03442323+2538451 & 0.997 & \\
   & BRB5 &  &  & 58.215833 & 23.563583 & 5174064 & J03525179+2333482 & 0.999 & \\
   & BRB6 &  &  & 58.29 & 23.563416 & 5173716 & J03530963+2333479 & 0.999 & \\
   & BRB7 &  &  & 57.18625 & 24.622972 & 5279607 & J03484469+2437236 & 1.0 & \\
   & BRB8 &  &  & 58.024166 & 24.292138 & 5172897 & J03520582+2417312 & 1.0 & \\
   & BRB10 &  &  & 57.312916 & 24.606222 & 5277273 & J03491512+2436225 & 1.0 & \\
   & BRB12 &  &  & 56.9125 & 24.606138 & 5238416 & J03473901+2436223 & 1.0 & \\
   & BRB13 &  &  & 58.802083 & 23.293888 & 5058757 & J03551261+2317373 & 0.0 & \\
   & BRB16 &  &  & 57.130833 & 24.577138 & 5277223 & J03483153+2434372 & 1.0 & \\
   & BRB17 &  &  & 58.534624 & 23.909277 &  &  &  & (1)\\
   & BRB19 &  &  & 58.714541 & 23.753388 & 5177022 & J03545147+2345122 & 0.0 & \\
   & BRB21 & PLZJ-4 &  & 58.542791 & 23.694527 & 5169447 & J03541027+2341402 & 1.0 & \\
   & BRB22 & PLZJ-61 &  & 56.130291 & 25.587527 & 5390123 & J03443127+2535149 & 1.0 & \\
   & BRB23 &  &  & 57.664708 & 25.048472 & 5278239 & J03503955+2502545 & 1.0 & \\
   & BRB24 &  &  & 57.081875 & 24.869305 &  &  &  & (1)\\
   & BRB25 &  &  & 58.248875 & 23.865555 & 5176839 & J03525971+2351557 & 0.0 & \\
   & BRB26 &  &  & 57.234041 & 25.161972 &  &  &  & (1)\\
   & BRB27 & PLZJ-32 &  & 56.113499 & 25.744972 & 5388791 & J03442728+2544419 & 0.996 & \\
   & BRB28 & PLZJ-37 &  & 58.228833 & 24.621833 & 5290218 & J03525491+2437181 & 0.999 & \\
   & BRB29 &  &  & 58.505958 & 23.832805 & 5171913 & J03540143+2349575 & 0.998 & \\
   & BRB30 &  &  & 56.987333 & 24.482888 &  &  &  & (1)\\
   & BRB31 &  &  & 57.858041 & 24.258944 &  &  &  & (1)\\
   & BRB32 &  &  & 57.253583 & 24.694027 &  &  &  & (1)\\
   & BRB33 &  &  & 57.833958 & 23.755111 & 2535230 & J03512013+2345178 & 0.0 & \\
   & BRB34 &  &  & 57.193958 & 24.750888 &  &  &  & (1)\\
   &  & PLZJ-56 &  & 56.222958 & 25.605405 & 5390145 & J03445352+2536194 & 0.0 & \\
   &  & PLZJ-45 &  & 58.2425 & 24.292102 &  &  &  & \\
   &  & PLZJ-50 &  & 55.983249 & 25.607069 & 4055668 & J03435600+2536253 & 1.0 & \\
   &  & PLZJ-60 &  & 56.134666 & 25.421683 & 5387838 & J03443232+2525181 & 1.0 & \\
   &  & PLZJ-46 &  & 58.238333 & 24.266944 &  &  &  & \\
   &  & PLZJ-77 &  & 58.541833 & 23.297855 &  &  &  & \\
   &  & PLZJ-10 &  & 58.129958 & 24.774891 &  &  &  & \\
   &  & PLZJ-23 &  & 57.972416 & 24.636697 &  &  &  & \\
   &  & PLZJ-93 &  & 58.804166 & 24.604388 & 2629741 & J03551301+2436158 & 0.0 & \\
   &  & PLZJ-323 &  & 55.980291 & 25.723944 & 2646887 & J03435527+2543262 & 0.0 & \\
   &  & PLZJ-721 &  & 58.779749 & 24.956205 & 2618690 & J03550717+2457230 & 0.044 & \\
   &  & PLZJ-235 &  & 58.135708 & 24.743511 &  &  &  & \\
   &  & PLZJ-112 &  & 58.330708 & 24.89218 & 2836575 & J03531938+2453317 & 0.0 & \\
   &  & PLZJ-100 &  & 56.829958 & 25.848138 &  &  &  & \\
   &  &  & PNM01 & 57.174958 & 22.916166 & 5046135 & J03484198+2254583 & 0.0 & \\
   &  &  & PNM02 & 55.461541 & 25.276361 & 5231515 & J03415076+2516351 & 0.0 & \\
   &  &  & PNM03 & 57.02925 & 22.944361 & 5006623 & J03480702+2256396 & 0.0 & \\
   &  &  & PNM04 & 57.130666 & 25.715861 & 5399403 & J03483135+2542571 & 0.0 & \\
   &  &  & PNM05 & 57.013291 & 23.201611 & 5005906 & J03480319+2312057 & 0.0 & \\
   &  &  & PNM06 & 55.525875 & 25.075277 & 5231041 & J03420620+2504310 & 0.0 & \\
   &  &  & PNM07 & 55.466583 & 25.216805 & 5231375 & J03415198+2513005 & 0.0 & \\
   &  &  & PNM08 & 57.134416 & 25.715833 & 5399402 & J03483226+2542569 & 0.0 & \\
   &  &  & PNM09 & 55.781 & 25.266638 & 5225601 & J03430744+2515598 & 0.0 & \\
   &  &  & PNM10 & 56.991666 & 25.820138 & 5393154 & J03475799+2549125 & 0.0 & \\
   &  &  & PNM11 & 57.016458 & 22.876999 & 5007559 & J03480394+2252372 & 0.0 & \\
   &  &  & PNM12 & 55.69625 & 25.199194 & 5225443 & J03424710+2511573 & 0.0 & \\
   &  &  & PNM13 & 55.789624 & 25.268361 & 4006442 & J03430951+2516060 & 0.0 & \\
   &  &  & PNM14 & 57.001333 & 25.690083 & 4025181 & J03480033+2541242 & 0.0 & \\
   &  &  & PNM15 & 57.162833 & 22.887749 & 4000699 & J03483909+2253158 & 0.0 & \\
   &  &  & Calar Pleiades  8 & 55.892458 & 25.037944 & 5226398 & J03433419+2502166 & 0.96 & \\
   &  &  & Calar Pleiades  9 & 55.512875 & 24.061416 & 5144919 & J03420308+2403413 & 0.0 & \\
   &  &  & Calar Pleiades 10 & 56.558583 & 23.365666 & 5008540 & J03461406+2321565 & 1.0 & \\
   &  &  & Calar Pleiades 11 & 55.575083 & 24.919444 & 5228413 & J03421802+2455099 & 1.0 & \\
   &  &  & Calar Pleiades 12 & 55.70075 & 24.066999 & 5139930 & J03424817+2404012 & 1.0 & \\
   &  &  & Calar Pleiades 13 & 55.433083 & 25.058472 & 5228741 & J03414395+2503305 & 1.0 & \\
   &  &  & Calar Pleiades 14 & 56.460958 & 23.743555 & 5155777 & J03455062+2344369 & 1.0 & \\
   &  &  & Calar Pleiades 15 & 55.627458 & 25.04425 & 5228710 & J03423058+2502392 & 1.0 & \\
   &  &  & Calar Pleiades 16 & 56.944916 & 25.587944 & 5394991 & J03474678+2535166 & 1.0 & \\
   &  &  & Calar Pleiades 17 & 56.298874 & 23.695444 & 5148223 & J03451172+2341436 & 1.0 & \\
   &  &  & Calar Pleiades 18 & 56.388749 & 23.576194 & 5153508 & J03453330+2334342 & 1.0 & \\
   &  &  & Calar Pleiades 19 & 57.065208 & 25.835777 & 5393192 & J03481565+2550089 & 0.999 & \\
   &  &  & Calar Pleiades 20 & 56.493666 & 23.698277 & 5153687 & J03455848+2341539 & 0.759 & \\
   &  &  & Calar Pleiades 21 & 56.7515 & 23.365777 &  &  &  & \\
   &  &  & Calar Pleiades 22 & 56.501625 & 23.326083 &  &  &  & \\
   &  &  & Calar Pleiades 23 & 55.729708 & 24.150722 &  &  &  & (2)\\
   &  &  & Calar Pleiades 24 & 56.702791 & 23.480194 &  &  &  & \\
   &  &  & Calar Pleiades 25 & 57.110916 & 22.864972 &  &  &  & \\
   &  &  & Calar Pleiades 26 & 56.317041 & 23.585361 & 5148179 & J03451608+2335088 & 0.0 & misidentification\\
   &  &  & Calar Pleiades 27 & 56.550541 & 23.304361 & 3387884 & J03461212+2318156 & 0.0 & \\
   &  &  & Calar Pleiades 28 & 56.657875 & 23.424249 &  &  &  & (2)\\
   &  &  & Calar Pleiades 29 & 55.72575 & 24.136527 &  &  &  & (2)\\
   &  &  & Calar Pleiades 30 & 57.006208 & 25.613666 &  &  &  & \\
   &  &  & Calar Pleiades 31 & 56.427041 & 23.730388 &  &  &  & (3) \\
   &  &  & R1 & 55.753 & 25.084611 & 3446081 & J03430072+2505046 & 0.0 & \\
   &  &  & R2 & 57.005416 & 23.136972 &  &  &  & (2)\\
   &  &  & R3 & 57.117416 & 23.209166 &  &  &  & (2)\\
   &  &  & R4 & 57.456708 & 22.898749 & 3430893 & J03494961+2253556 & 0.0020 & \\
   &  &  & R5 & 57.076333 & 25.456305 & 2770428 & J03481832+2527227 & 0.0 & \\
\hline
\end{longtable}
\tablefoot{(1) No detection even though the survey was sensitive enough to detect the source; (2) $i$-band luminosity inconsistent with membership; (3) $r$-band luminosity inconsistent with membership;  }
\end{longtab}

\subsection{\citet{2006A&A...458..805B}}
\citet{2006A&A...458..805B} complemented the CFH12K data presented in \citet{2003A&A...400..891M} with deep near-infrared J-band images to search for new substellar members. A total of 26 out of 34 very low mass and substellar members of their list have a counterpart in the DANCe catalog. None of the eight that are missing are detected in our deeper MegaCam stacked images even though they are supposed to be brighter than the limit of detection, so we consider them as false detections. Finally, four of the twenty-six candidates with a counterpart in the DANCe catalog have membership probabilities close to zero essentially because of inconsistent proper motions, and are also considered as contaminants. The remaining 22 have high membership probabilities($>$99\%) and are considered as good candidate members. Table~\ref{tab:comparison} gives a summary of the 12 candidate members from \citet{2006A&A...458..805B}  rejected in our analysis.

\subsection{\citet{2007MNRAS.378.1131C}}
\citet{2007MNRAS.378.1131C} combined the UKIDSS ZYJHK images with the far red $I,Z$ CFH12K images presented in  \citet{2003A&A...400..891M} to search for new members. A total of 16 out of 23 candidate members have a counterpart in the DANCe catalog. Five out of these 16 candidate members have very low membership probabilities. All but one have proper motions that are inconsistent with the cluster's motion. The last one (PLZJ-56) has a proper motion consistent with the cluster's mean motion but optical luminosities and colors inconsistent with the cluster's sequence: the source appears to be slightly bluer (by $\sim$0.2~mag) than the Pleiades sequence in all color-magnitude diagrams. This object could be a genuine member with a peculiar color, so it deserves further attention. Follow-up observations are needed to confirm its membership status. The remaining 11 have high membership probabilities($>$99\%) and are considered to be good candidate members. Table~\ref{tab:comparison} gives a summary of the 16 candidate members from \citet{2007MNRAS.378.1131C} with a counterpart in the DANCe survey.

\subsection{\citet{2014A&A...568A..77Z}}

\citet{2014A&A...568A..77Z} have recently reported the discovery of 44 ultracool dwarf candidate members of the Pleiades based on proper motions and near-infrared photometry.
Thirty two sources from their Tables 1 and 2 have a counterpart in the DANCe catalog \citep[][ and Table~\ref{tab:comparison}]{2014A&A...563A..45S}. Twenty have a membership probability close to 0 because their proper motions are inconsistent with the cluster's mean motion within the uncertainties. 

\begin{figure}
   \centering
   \includegraphics[width=0.45\textwidth]{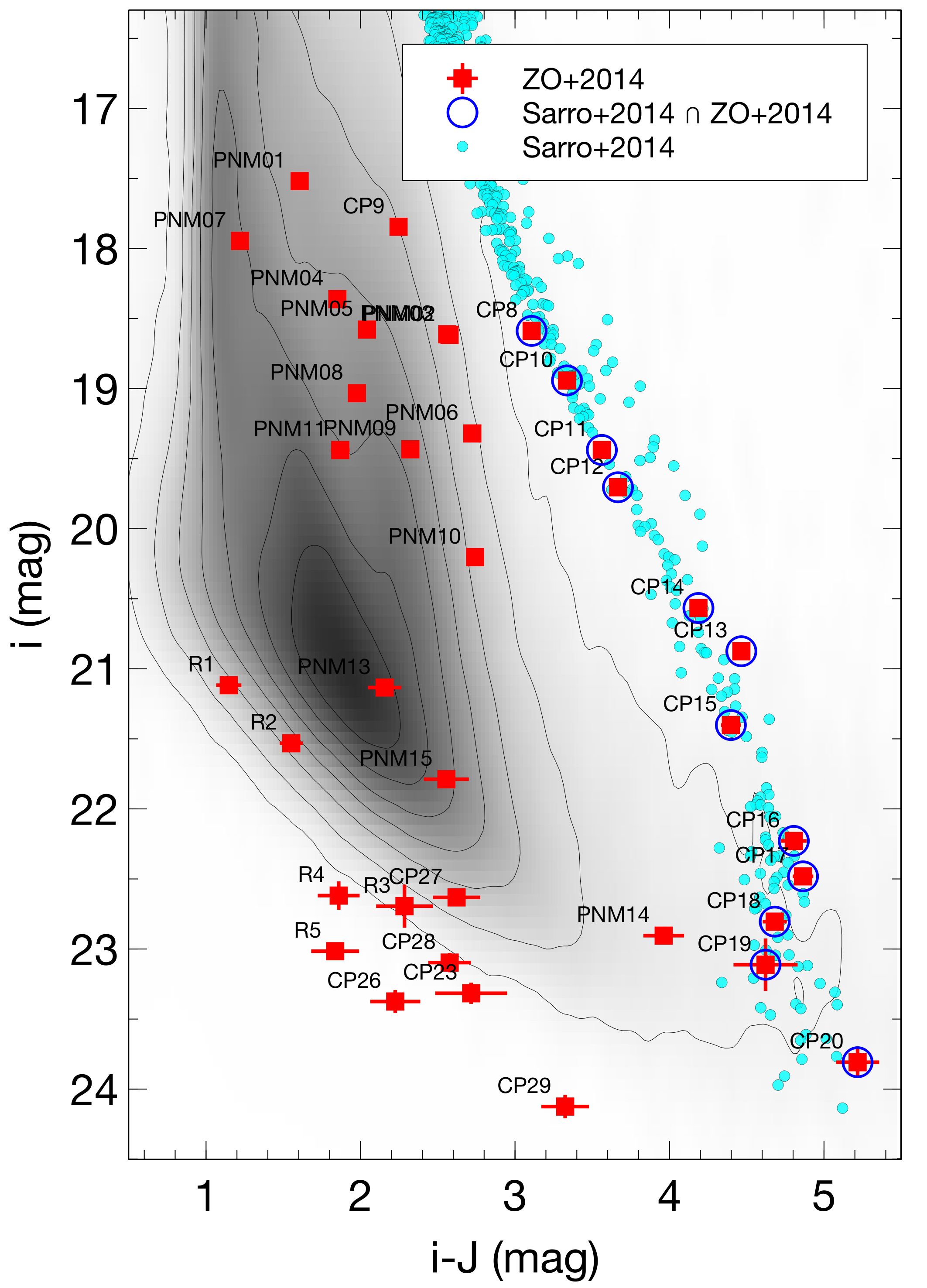}
      \caption{$i$ vs $i-J$ color-magnitude diagram of the DANCe survey. High probability ($\ge$0.95) sources are indicated as cyan dots. Candidate members from \citet{2014A&A...568A..77Z} with a $i$-band counterpart in the DANCe survey are represented with red squares. Their abbreviated names are also indicated. Candidate members from \citet{2014A&A...568A..77Z} previously discovered or reported in \citet{2014A&A...563A..45S}  are overplotted as large blue circles.}
         \label{fig:ZO_iJ}
   \end{figure}

In most cases, the DANCe dataset used to derive these proper motions are excellent, with a long time baseline, a relatively large number of epochs, and reduced-$\chi^{2}$ close to unity. The deep multi-wavelength photometry included in the DANCe catalog is also useful for testing their membership. Several are detected in either $u$, $g,$ or $r$ or in the old DSS plates, which is highly unlikely for such ultracool Pleiades brown dwarfs. All are detected in either the $r$ or $i$-band with luminosities inconsistent with the empirical Pleiades sequence, as shown in Fig.~\ref{fig:ZO_iJ}. On the basis of our more precise proper motion measurements, optical photometry, and probabilistic membership analysis, we rejected these 20 candidates and classified them as contaminants. Only 12 candidates of the 32 with a counterpart in the DANCe catalog have high membership probabilities according to \citet{2014A&A...563A..45S}. All 12 had been previously discovered and identified in the literature or in the DANCe survey. 

Finally, half of the 12 sources without counterparts in  \citet{2014A&A...563A..45S} have a counterpart in the complete DANCe catalog but no reliable proper motion, so they were not included in the \citet{2014A&A...563A..45S} analysis and catalog. The DANCe complete catalog includes photometric information that helps discard a few more candidates. Calar Pleiades 31 is detected in the $r$ band with 22.66$\pm$0.12~mag -- too bright for a J=21.37~mag Pleiades ultracool dwarf. Calar Pleiades 23, 28, 29 and R2, R3 are detected in the $i$-band with luminosities inconsistent with the empirical Pleiades sequence (see Fig.~\ref{fig:ZO_iJ}). Calar Pleiades 22, 24, 25, 26, 30, and 31 do not have any counterpart in the DANCe complete catalog. According to the authors, Calar Pleiades 26 is associated to a faint source in the UKIDSS $Ks$ image, which they use to derive a proper motion measurement. A visual inspection of the UKIDSS and WISE images shows that the cross-identification is wrong, because the counterpart is located 1\farcs4 north of the reported position of Calar Pleiades 26 and is associated to an optical source (DANCe-J03451608+2335088, within 0\farcs2). The corresponding astrometry and photometry is therefore wrong. The detection seen near (but not exactly at) the location of Calar Pleiades 26 in the same Omega-Prime images as the one used by these authors is extremely faint and, if real, is blended with the DANCe source 5148179. For the rest of the study, we consider Calar Pleiades 26 as a contaminant. A total of five candidates remain, leading to a contamination rate of $\ge$61\% in the \citet{2014A&A...568A..77Z} survey.  

From this original sample, \citet{2014A&A...572A..67Z} have recently obtained near-infrared spectra of the candidate members DANCe-J03474678+2535166, DANCe-J03453330+2334342, DANCe-J03481565+2550089, DANCe-J03455848+2341539, and Calar Pleiades 21 and 22. The spectroscopic features support their cluster membership. They also obtained an improved upper limit on the $Z$-band photometry of Calar Pleiades 25, which is consistent with membership but does not allow drawing a firm conclusion about its membership. 

\subsection{Importance of inspecting the images}
Proper motion surveys based on a couple of epochs have been commonly used to search for members of young clusters. The present analysis finds contamination rates of at least 50\%, 35\%, 43\%, and 61\% in \citet{2003A&A...400..891M}, \citet{2006A&A...458..805B}, \citet{2007MNRAS.378.1131C}, and \citet{2014A&A...568A..77Z}, respectively. The vast majority of identified contaminants have discrepant proper motions according to the DANCe measurements, which illustrates the need to have more than just two epochs to derive robust proper motions from ground-based images. 

Additionally, a significant fraction of the contamination is related to false detections and illustrates the need to always visually inspect the images and not rely just on the catalogs. The latter inevitably contain errors and false or problematic detections. These recommendations also apply to the DANCe catalog itself, even though it is based on a larger number of images and epochs (so is generally more reliable). For that reason, we make the MegaCam $r,i$ and UKIDSS $Ks$ mosaics used for the DANCe survey available on the internet at the following address: \url{http://visiomatic.iap.fr/pleiades/}. The images can be browsed in an intuitive, interactive way thanks to the VisiOmatic tool \citep{2014arXiv1403.6025B}. The Pleiades members identified in the DANCe survey are indicated, and their photometry, spectral energy distribution (SED), best-fit effective temperatures (see Section~\ref{sec:lumteff}), proper motions, and membership probabilities are reported.

\section{Empirical Pleiades isochrones \label{sec:isochrones}}

\subsection{Empirical isochrones}
\begin{figure*}
   \centering
   \includegraphics[height=0.95\textheight]{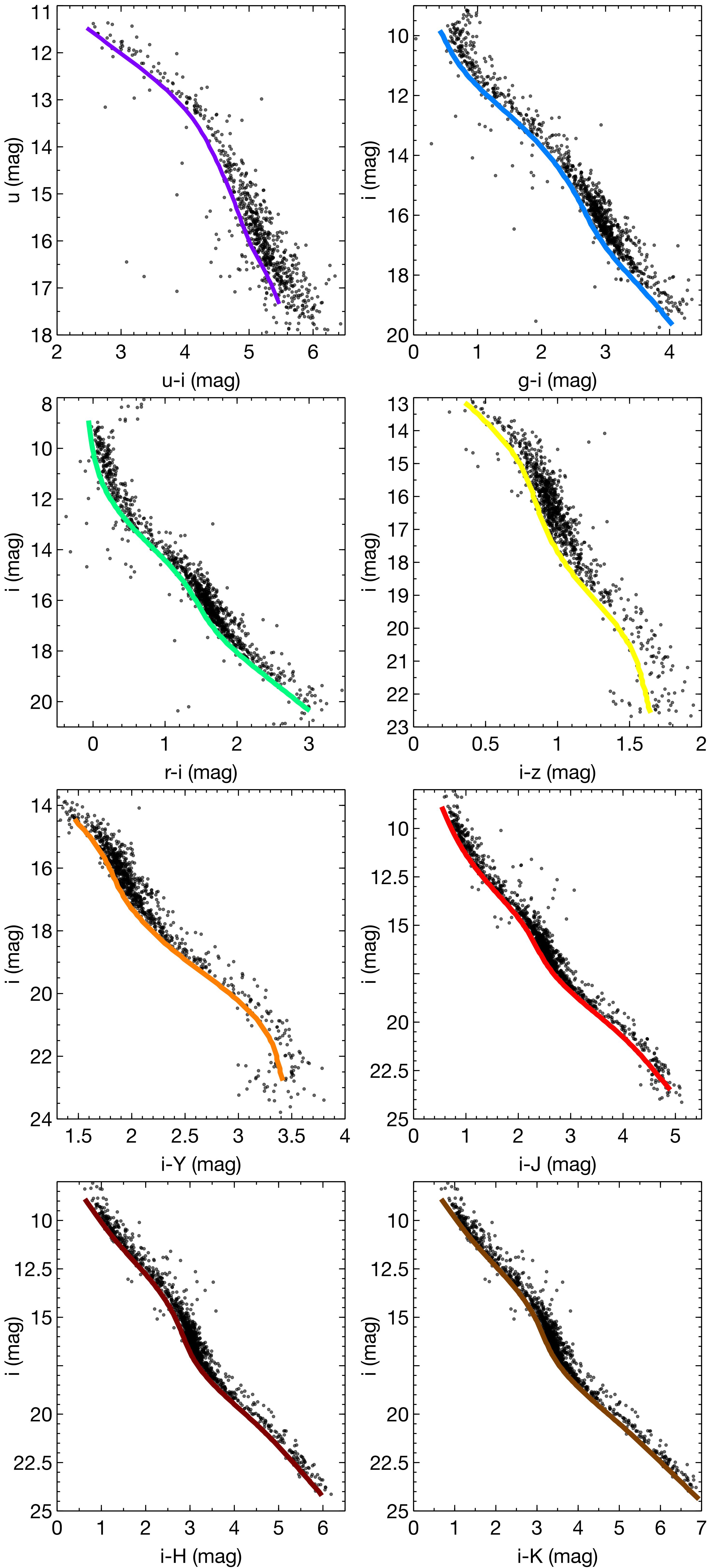}
      \caption{Empirical isochrones in the $u$,$g$,$r$,$i$,$Y$,$J$,$H$,$Ks$ using the $i$-band as reference (colored curves) overplotted on top of high probability members (proba$\ge$0.99). }
         \label{fig:isochrones}
   \end{figure*}

The robust membership probabilities and multiwavelength photometry covering the spectral range from the near-UV the near-infrared offers a unique opportunity to measure precise empirical isochrones. Such isochrones prove extremely useful as a reference for the study of other groups and clusters and can be used as an absolute empirical calibration for the evolutionary models. Table~\ref{tab:isochrones-dance} gives the apparent magnitudes for the Pleiades empirical sequence in the $u$,$g$,$r$,$i$,$Y$,$J$,$H$,$Ks$ over the dynamic range of the DANCe dataset. The $i$-band was used as reference, and the other luminosities were imputed by fitting a principal curve to the ridge of the empirical sequence in the corresponding color-magnitude diagram and by applying a manual offset (typically between 0.1 and 0.2~mag in color) to match the lower edge, as illustrated in Figure~\ref{fig:isochrones}. To minimize systematics and errors, only sources with a membership probability greater than or equal to 0.99 were used. Similarly, Table~\ref{tab:isochrones-tycho} gives the apparent magnitudes for the Pleiades empirical sequence in the $B_{\rm T}$, $V_{\rm T}$, J, H, and K-bands obtained from the reanalysis of the Tycho-2 data and using $V_{\rm T}$ as reference. Increasing photometric errors towards the faint end result in larger uncertainties for the location of the sequence. The isochrones should therefore be considered and used with caution at the faint end (see Fig~\ref{fig:isochrones}). 

\begin{table}
\scriptsize
\caption{Empirical isochrones of the Pleiades. More detailed version
  available on-line \label{tab:isochrones-dance}}
\begin{tabular}{rrrrrrrrr}
\hline
  \multicolumn{1}{c}{i} &
  \multicolumn{1}{c}{u} &
  \multicolumn{1}{c}{g} &
  \multicolumn{1}{c}{r} &
  \multicolumn{1}{c}{z} &
  \multicolumn{1}{c}{Y} &
  \multicolumn{1}{c}{J} &
  \multicolumn{1}{c}{H} &
  \multicolumn{1}{c}{Ks} \\
\hline
  9.4 &  &  & 9.36 &  &  & 8.779 & 8.617 & 8.556\\
  9.9 &  & 10.33 & 9.886 &  &  & 9.195 & 8.962 & 8.886\\
  10.4 &  & 10.95 & 10.42 &  &  & 9.604 & 9.301 & 9.209\\
  10.9 &  & 11.59 & 10.97 &  &  & 10.0 & 9.629 & 9.522\\
  11.4 &  & 12.27 & 11.53 &  &  & 10.39 & 9.942 & 9.821\\
  11.9 & 14.78 & 13.0 & 12.12 &  &  & 10.75 & 10.25 & 10.11\\
  12.4 & 15.77 & 13.76 & 12.74 &  &  & 11.11 & 10.55 & 10.39\\
  12.9 & 16.69 & 14.52 & 13.38 &  &  & 11.45 & 10.86 & 10.67\\
  13.4 & 17.52 & 15.26 & 14.04 & 12.98 &  & 11.79 & 11.18 & 10.97\\
  13.9 & 18.25 & 15.97 & 14.72 & 13.36 &  & 12.12 & 11.52 & 11.28\\
  14.4 & 18.93 & 16.66 & 15.39 & 13.76 &  & 12.46 & 11.88 & 11.62\\
  14.9 & 19.59 & 17.32 & 16.05 & 14.18 & 13.32 & 12.82 & 12.26 & 11.98\\
  15.4 & 20.23 & 17.96 & 16.69 & 14.62 & 13.71 & 13.2 & 12.66 & 12.36\\
  15.9 & 20.87 & 18.58 & 17.3 & 15.07 & 14.12 & 13.6 & 13.07 & 12.76\\
  16.4 & 21.55 & 19.2 & 17.91 & 15.53 & 14.55 & 14.0 & 13.48 & 13.15\\
  16.9 & 22.24 & 19.85 & 18.52 & 15.99 & 14.98 & 14.39 & 13.88 & 13.54\\
  17.4 &  & 20.52 & 19.16 & 16.44 & 15.37 & 14.76 & 14.25 & 13.89\\
  17.9 &  & 21.23 & 19.84 & 16.87 & 15.74 & 15.1 & 14.59 & 14.21\\
  18.4 &  & 21.94 & 20.54 & 17.29 & 16.09 & 15.41 & 14.9 & 14.49\\
  18.9 &  & 22.65 & 21.26 & 17.69 & 16.41 & 15.71 & 15.18 & 14.75\\
  19.4 &  & 23.34 & 21.98 & 18.09 & 16.71 & 15.99 & 15.44 & 14.99\\
  19.9 &  &  & 22.7 & 18.49 & 17.02 & 16.27 & 15.7 & 15.22\\
  20.4 &  &  &  & 18.92 & 17.35 & 16.55 & 15.96 & 15.45\\
  20.9 &  &  &  & 19.36 & 17.71 & 16.85 & 16.23 & 15.69\\
  21.4 &  &  &  & 19.82 & 18.11 & 17.17 & 16.51 & 15.93\\
  21.9 &  &  &  & 20.29 & 18.55 & 17.5 & 16.81 & 16.18\\
  22.4 &  &  &  & 20.77 & 19.01 & 17.84 & 17.11 & 16.43\\
  22.9 &  &  &  &  &  & 18.19 & 17.42 & 16.69\\
  23.4 &  &  &  &  &  & 18.53 & 17.72 & 16.95\\
  23.9 &  &  &  &  &  &  & 18.03 & 17.21\\
\hline\end{tabular}
\end{table}

\begin{table}
\tiny
\caption{Empirical isochrones of the Pleiades in the $B_{\rm
    T}$,$V_{\rm T}$, J, H and Ks bands. More detailed version
  available on-line \label{tab:isochrones-tycho}}
\begin{tabular}{rrrrr}
\hline
  \multicolumn{1}{c}{$V_{\rm T}$} &
  \multicolumn{1}{c}{$B_{\rm T}$} &
  \multicolumn{1}{c}{J} &
  \multicolumn{1}{c}{H} &
  \multicolumn{1}{c}{Ks} \\
\hline
  3.9 &  & 4.09 &  & 4.076\\
  4.4 & 4.329 & 4.517 & 4.591 & 4.532\\
  4.9 & 4.831 & 4.978 & 5.054 & 5.002\\
  5.4 & 5.339 & 5.444 & 5.514 & 5.47\\
  5.9 & 5.862 & 5.897 & 5.957 & 5.926\\
  6.4 & 6.396 & 6.329 & 6.377 & 6.35\\
  6.9 & 6.946 & 6.741 & 6.768 & 6.739\\
  7.4 & 7.529 & 7.119 & 7.113 & 7.086\\
  7.9 & 8.118 & 7.466 & 7.416 & 7.39\\
  8.4 & 8.716 & 7.775 & 7.691 & 7.65\\
  8.9 & 9.315 & 8.088 & 7.962 & 7.906\\
  9.4 & 9.925 & 8.423 & 8.248 & 8.185\\
  9.9 & 10.56 & 8.779 & 8.561 & 8.492\\
  10.4 &  & 9.074 & 8.871 & 8.795\\
  10.9 &  & 9.252 & 9.143 & 9.071\\
  11.4 &  &  & 9.287 & 9.239\\
\hline\end{tabular}
\end{table}

\subsection{Comparison with evolutionary models \label{sec:compmodel}}
\begin{figure*}
   \centering
   \includegraphics[height=0.95\textheight]{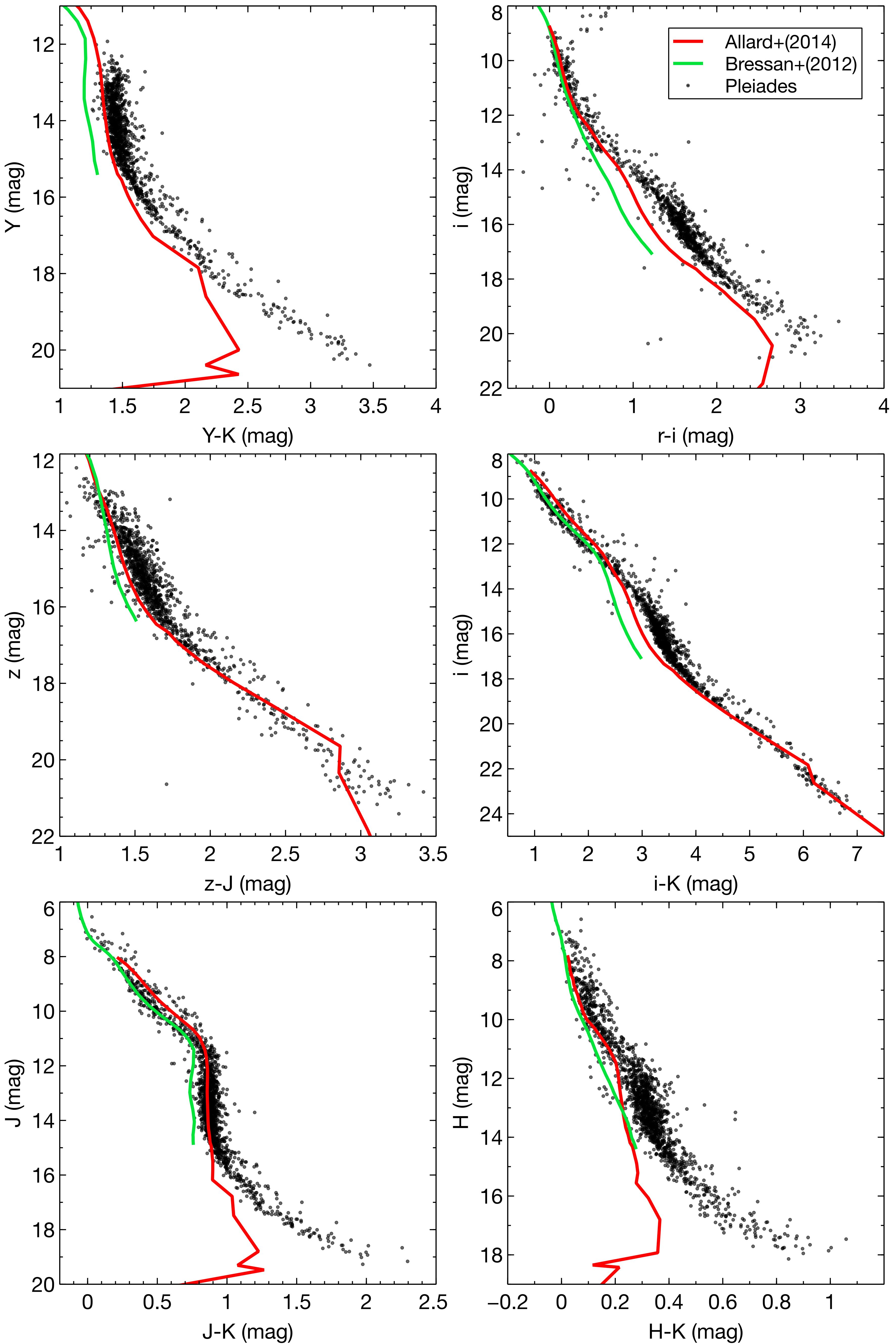}
      \caption{Comparison of the observations of the Pleiades sequence (black dots) and models from \citet{2014IAUS..299..271A} (red) and \citet{2012MNRAS.427..127B} (light green) for 120~Myr in various color magnitude diagrams. }
         \label{fig:comparison}
   \end{figure*}

Figure~\ref{fig:comparison} compares the observed sequence of Pleiades members in various color-magnitude diagrams to the models of \citet{2014IAUS..299..271A} and \citet{2012MNRAS.427..127B} for an age of 120~Myr. Members with a probability greater or equal to 0.99 are represented to minimize the contamination and associated noise in the comparison.
The match is in general relatively poor in any color-magnitude diagram made of optical and near-infrared luminosities.  As already known \citep[see, e.g.,][ and references therein]{2012MNRAS.424.3178B,1998A&A...337..403B}, the difference between observations and models is larger at the faint end and suggests that the current description of the atmospheres of late-M and L-type brown dwarfs presents a number of problems. Surface inhomogeneity \citep[active zones, spots, ][]{2003AJ....126..833S, 2014AJ....148...30K}, and rotation \citep{2014arXiv1410.4238S,2014AJ....148...30K}, molecular opacity \citep[e.g.,][]{2012ApJ...752...56F,2014A&A...564A..55M}, weather \citep{2014Natur.505..654C}, and abundances \citep[specifically the solar abundances as reference, see, e.g.,][]{2011ASPC..448...91A} are all known to contribute to the differences between observations and model predictions. Such models should therefore be used with caution in the corresponding luminosity range. Whenever possible and until the models match the observations, empirical isochrones, such as the ones provided above should be preferred to their theoretical counterparts when looking for members.

\section{Present-day luminosity and mass functions of the Pleiades cluster  \label{sec:pdmf}}

\subsection{Luminosities and effective temperatures of the Pleiades members \label{sec:lumteff}}
We estimated the effective temperatures (hereafter T$_{\rm eff}$) and bolometric luminosity (L$_{\rm bol}$) of all Pleiades candidate members using the {\it virtual observatory SED analyzer} \citep[hereafter VOSA, ][]{2008A&A...492..277B}. For that purpose, the optical and near-infrared photometry of the DANCe catalog was complemented by {\it WISE} \citep{2012wise.rept....1C} mid-infrared photometry, when available. 

Briefly, VOSA compares the observed SED  to a grid of  theoretical SEDs. The grid was made of BT-Settl models \citep{2012EAS....57....3A}  in the range 1\,500$<$T$_{\rm eff}<$4\,000~K (in steps of 100~K) and Kurucz models \citep{Kurucz} in the range T$_{\rm eff}>$3\,500~K. The overlap was defined on purpose to mitigate the effect of the discontinuity between the two sets of models over that range of effective temperatures. In that range, the best fit (in terms of $\chi^{2}$) was chosen between the two sets of models.

VOSA offers the advantage of deriving L$_{\rm bol}$ using all the available photometric information rather than a subset of colors and luminosities. It estimates the bolometric luminosities by integrating the observed luminosities complemented by the best-fit model over the wavelength range not covered by the observations. For most of the sample, at least 35\% of the inferred bolometric luminosity corresponds to the observations. As such, it is less (but nevertheless still) sensitive to problems in a given photometric band pass, uncertain bolometric corrections, errors in the evolutionary models as described previously, variability, excesses, and some extent extinction. The T$_{\rm eff}$ is computed using the L$_{\rm bol}$ derived as described above and assuming a stellar radius. The latter depends  mostly  on the interior models, which have been relatively well calibrated for low mass stars. The T$_{\rm eff}$ should therefore be only moderately affected by the deficiencies in the atmospheric models described in Section~\ref{sec:compmodel}. Using a similar method, \citet{2002AJ....124.1170D} estimated errors on the T$_{\rm eff}$  on the order of 50~K, which is smaller than the sampling of the grid of models used in our study.

The photometry was deredenned by assuming the canonical mean extinction to the Pleiades of A$_{\rm V}$=0.12~mag and a distance of 136.2~pc \citep{2014Sci...345.1029M} was assumed to compute the bolometric luminosities. Photometric measurements below 0.45~$\mu$m and beyond 12~$\mu$m are often subject to excesses related to activity and/or the presence of circumstellar debris disks and were systematically ignored for the fit. A visual inspection of the SEDs was performed to reject the remaining obviously problematic photometric measurements. The quality and the quantity of the photometric measurements ($\sim$85\% of the sources have more than 8 photometric measurements) resulted in a robust fit for most sources. Only six sources had fewer than the minimum three photometric measurements required to perform a meaningful fit, and are discarded for the rest of the analysis. The DANCe Pleiades Release 2 online table includes the estimated T$_{\rm eff}$ and L$_{\rm bol}$ for the 2\,010 sources selected as members in this  study.

\subsection{Completeness and contamination of the sample\label{sec:completeness}}
Prior to deriving a luminosity function, we assess the completeness of our analysis. We identify several sources of incompleteness and contamination. As described in \citet{2014A&A...563A..45S}, the 75\% selection threshold on the membership probability leads to an estimated contamination rate of $\sim$7\%  at most. The number of missed members is estimated to reach 4\% at most. These numbers do not take real Pleiades members with properties differing significantly from their siblings into account:
\begin{itemize}
\item High order ($N\ge$3) unresolved multiple systems can appear significantly brighter than the single star and equal-mass binary sequences used for the calculations and hence be given lower probabilities. Figure~\ref{fig:i_iK} shows that several candidate members passed our selection criterion in spite of being overluminous and could be high order multiple systems.
\item The orbital motion of long period (decades or centuries) unresolved binaries can affect their apparent proper motion significantly and result in lower membership probabilities.
\item White dwarfs have colors and luminosities inconsistent with the empirical stellar main sequences used to compute the membership probabilities and were missed. A future study will look for them in the DANCe catalog.
\end{itemize}

Additionally, because the DANCe catalog is based on archival observations of very diverse origins, the sensitivity -- hence completeness -- is not homogeneous over the entire area covered by the survey. Based on the spatial coverage and depth of the datasets used to derive the DANCe catalog, we identify a region of the survey where the completeness is believed to be homogeneous. This area is defined by our own recent CFHT MegaCam $i$-band observations and covers a roughly square region of $\sim$3$\times$3\degr\, centered on the Seven Sisters \citep[see Fig.~1 of ][]{2013A&A...554A.101B}. To perform a meaningful study of the luminosity and mass functions, we therefore limit the following analysis to a circular region around the cluster's center (at RA=03:46:48 and Dec=24:10:17 J2000) and with a radius of 3\degr. A total of 1\,378 members fall within this region. 

\citet{1998MNRAS.299..955P} reported a significant mass segregation that might affect the present study of the mass and luminosity function. We nevertheless note that the 3\degr\, radius corresponds to more than double the size of the core radius for low mass stars and almost seven times the core radius for high mass stars estimated by \citet{1998MNRAS.299..955P}. 

\subsection{System's present-day luminosity function}

\begin{figure*}
   \centering
   \includegraphics[width=0.95\textwidth]{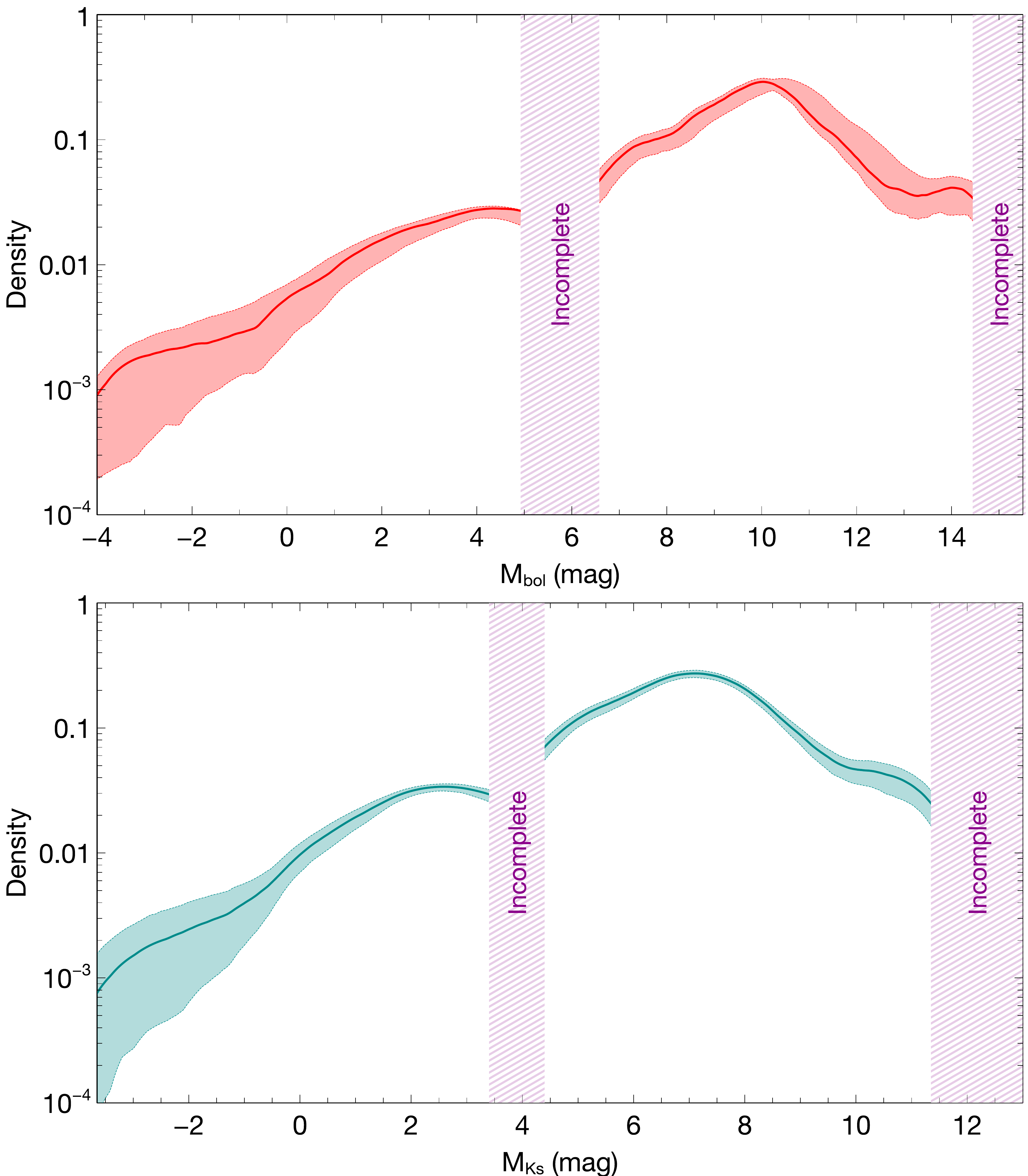}
      \caption{Top panel: Observed present-day bolometric luminosity function for systems within the central 3\degr\, of the Pleiades cluster. Bottom panel: Observed present-day absolute Ks-band luminosity function for systems  within the central 3\degr\, of the Pleiades cluster. Domains where the survey is incomplete are indicated by shaded areas.}
         \label{fig:lumfunc}
   \end{figure*}

Figure~\ref{fig:lumfunc} shows the bolometric and absolute Ks-band system's luminosity function assuming a distance of  136.2~pc. We chose to represent the distribution of absolute Ks-band magnitudes because all but 8 of the 1\,378 members falling inside the region defined above have a Ks-band measurement.The DANCe catalog is believed to be sensitive up to Ks$\sim$20~mag but complete up to Ks$\sim$17~mag (or $M_{\rm Ks}=$11.3~mag). A kernel density estimate (KDE) was preferred over histograms to build a continuous present-day luminosity function (PDLF). The KDE was obtained using a bandwidth of 0.3~mag for the DANCe sample and 0.8~mag for the Tycho-2 sample and an Epanechnikov kernel. Uncertainties were estimated using bootstrapping and computing the 95\% confidence intervals (defined between quantile 2.5\% and 97.5\%) over 1\,000 replicates. These uncertainties do not include the contamination and missed members mentioned above or the photometric uncertainties.

An interesting feature is observed beyond $M_{\rm Ks}\ge$9.8~mag (Ks$\ge$15.5~mag) and $M_{\rm bol}\ge$12.5~mag, where the luminosity function becomes almost flat. These luminosities correspond roughly to 0.04$\sim$0.05~M$_{\sun}$ according to the models of \citet{2014IAUS..299..271A} assuming an age of 120~Myr \citep{1998ApJ...499L.199S,Barrado_in_prep}. This significant change in the slope suggests that different or additional formation mechanisms must be at work for high and low mass brown dwarfs. 

We note that the small drop at the edges of the complete domain is interpreted as a boundary effect of the kernel density estimate and is not physical.
The luminosity function is available in the online Table~\ref{tab:bollumfunc}.

\subsection{System's present-day mass function}
Deriving a mass function requires transforming luminosities to masses. These transformations are usually based on poorly or uncalibrated theoretical mass-luminosity relationships. In Section~\ref{sec:isochrones}, we found that the models poorly match the observed colors and luminosities, especially at the very low mass end. The mass-luminosity relationship in the corresponding luminosity range must therefore be considered with caution, and any mass estimate should be regarded as strongly model dependent and tentative. With this important limitation in mind, we derive the Pleiades present-day mass function for system (PDMF) and compare it to recent predictions and models of the system's initial mass function (IMF).

The values of L$_{\rm bol}$ estimated with VOSA were transformed into masses using the mass-luminosity relationship given by the 120~Myr BT-Settl models for objects with T$_{\rm eff}\le$4\,000~K and the 120~Myr \citet{2012MNRAS.427..127B} models for objects hotter than T$_{\rm eff}\ge$4\,000~K. According to these calculations, the faintest candidate members have masses of only $\sim$20~M$_{\rm Jup}$, and the most massive member is Alcyone with 9.4~M$_{\sun}$. The estimated masses are also included in the online DANCe Pleiades Release 2 table.

\begin{figure*}
   \centering
   \includegraphics[width=0.95\textwidth]{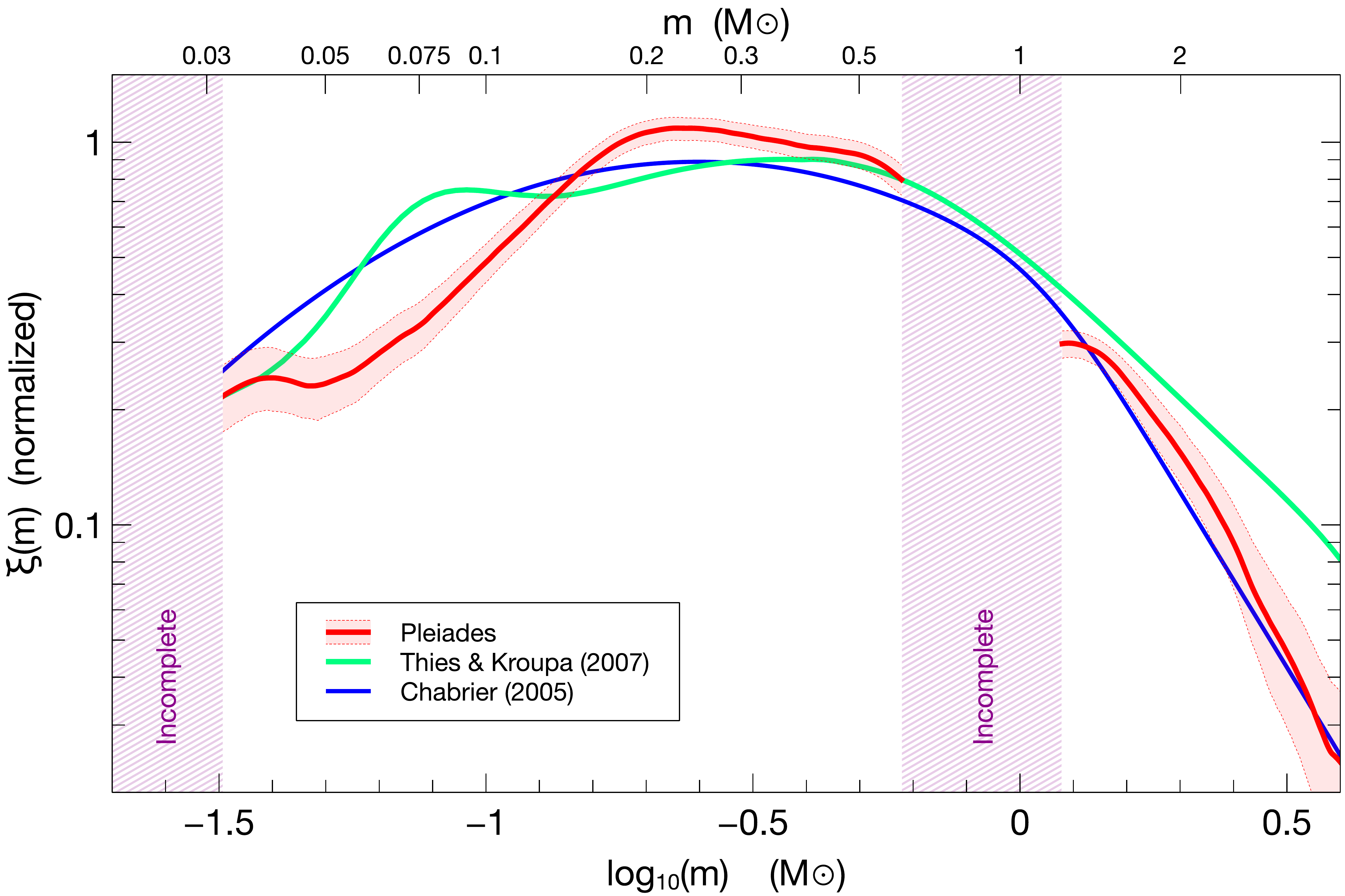}
      \caption{Observed present-day system mass function of the Pleiades within the central 3\degr\, radius (red). \citet{2005ASSL..327...41C} model for the galactic disk mass function (normalized to the same total mass over the range 0.03--0.6~M$_{\sun}$) and \citet{2007ApJ...671..767T} prediction for the Pleiades cluster are overplotted in blue and green, respectively, and normalized to the same total mass over the range 0.03--0.6~M$_{\sun}$.  The ``gap'' between the Tycho-2 and DANCe analyses and the estimated limit of completeness of the DANCe survey are indicated by shaded areas.}
         \label{fig:massfunc}
   \end{figure*}

Figure~\ref{fig:massfunc} shows the distribution of masses obtained. The KDE was obtained using a bandwidth of 0.1 (in $log_{10}(m)$) for the DANCe and Tycho-2 samples and an Epanechnikov kernel in both cases. As for the PDLF, uncertainties were estimated using bootstrapping and computing the 95\% confidence intervals (defined between quantiles 2.5\% and 97.5\%) over 1\,000 replicates, and they do not include the contamination and missed members mentioned in Section~\ref{sec:completeness}. Table~\ref{tab:massfunc} gives the results, and Fig.~\ref{fig:massfunc} compares to \citet{2005ASSL..327...41C} model and \citet{2007ApJ...671..767T} prediction normalized to the same total mass over the range 0.03--0.6~M$_{\sun}$.  Here as well the small drop at the edges of the complete domain is interpreted as a boundary effect of the kernel density estimate so is not physical.

\vspace{1cm}

\subsection{Comparison to theoretical predictions}
Unlike what was reported by previous authors \citep[see, e.g.,][]{2003A&A...400..891M, 2013pss5.book..115K}, the slope above 1~M$_{\sun}$ matches the $\alpha$=2.3 value predicted by  \citet{2007ApJ...671..767T} relatively well. But as mentioned above, the study of membership by \citet{2014A&A...563A..45S} used for the present analysis did not include white dwarfs. At least two white dwarf candidate members of the Pleiades were reported in the literature \citep[][]{2006MNRAS.373L..45D, Luyten1960}, one of them ultra-massive. If confirmed, these two members could significantly change the slope of the PDMF at the high mass end. Additionally, early dynamical interactions in multiple systems probably ejected a substantial number of massive stars. The missing white dwarfs and ejected members therefore affect the comparison with the model of \citet{2005ASSL..327...41C} and \citet{2007ApJ...671..767T}, and suggest that in spite of the apparently better match of  \citet{2005ASSL..327...41C} model with our observations, \citet{2007ApJ...671..767T} prediction is probably closer to the real distribution (in that mass range).

Overall, \citet{2005ASSL..327...41C} model and \citet{2007ApJ...671..767T}  match the PDMF in the range 0.02$<$M$<$0.6~M$_{\sun}$ relatively well. Both predict too many very low mass stars and brown dwarfs, a discrepancy that has been repeatedly reported in the Pleiades and other clusters and could be related to uncertainties in the mass-luminosity relationship used to build the observed PDMF. 
Interestingly, the PDMF becomes flat at the very end, between 20~M$_{\rm
Jup}$ and 50~M$_{\rm Jup}$, a feature already suggested by other authors and now confirmed by our study and the consequence of the flattening observed in the luminosity function mentioned above. One possible explanation was that several formation mechanisms lead to the formation of brown dwarfs. While the majority of brown dwarfs probably form like their more massive stellar counterparts from the fragmentation and collapse of a molecular cloud, a significant number might be formed as companions in the disk of a more massive system and subsequently ejected by dynamical interactions.

\begin{figure*}
   \centering
   \includegraphics[width=0.95\textwidth]{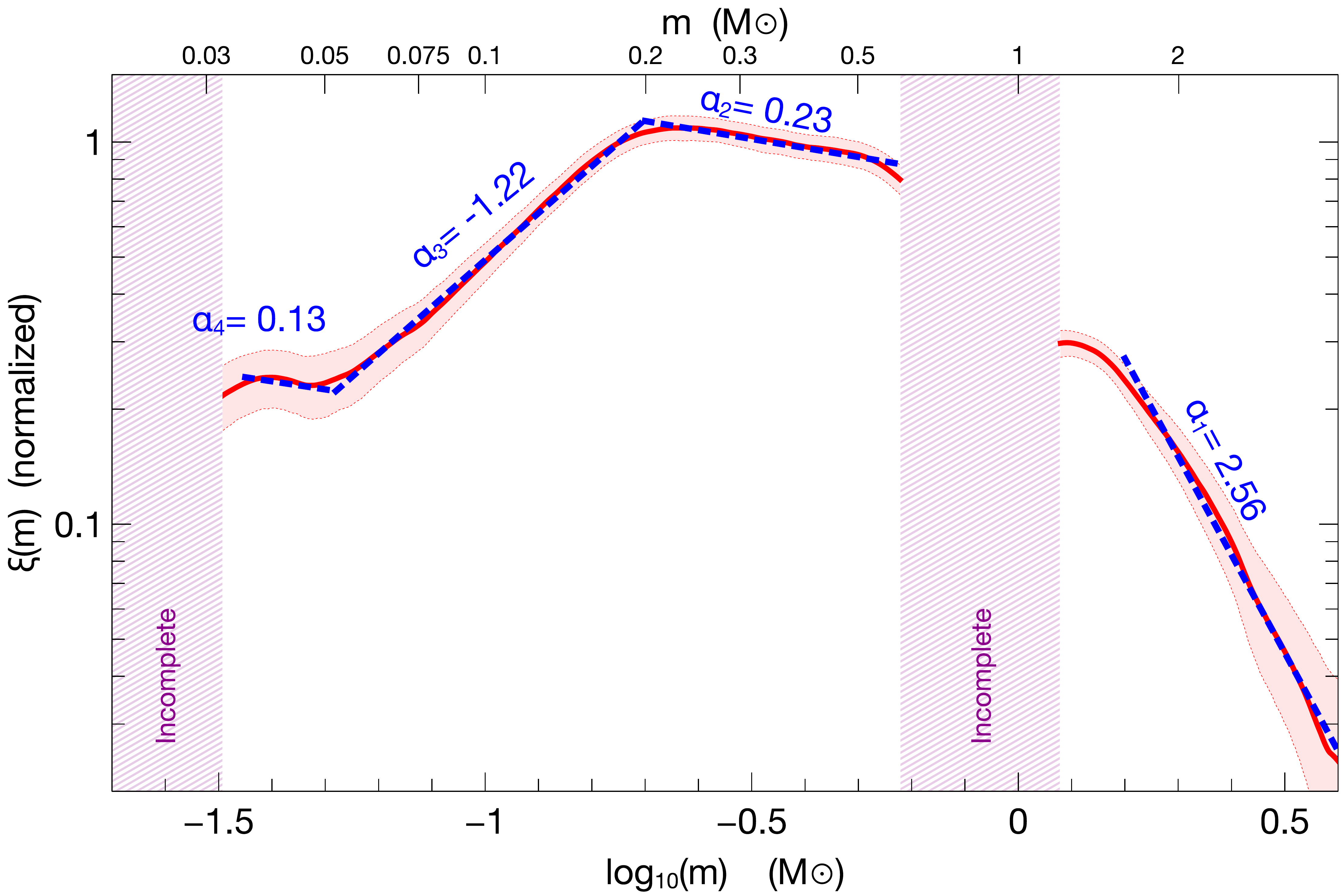}
      \caption{Observed present-day system mass function of the Pleiades within the central 3\degr\, radius (red). Best fit power laws in the form $m^{-\alpha}$  are overplotted in blue, and their index is indicated. The ``gap'' between the Tycho-2 and DANCe analyses is indicated by shaded areas.}
         \label{fig:massfunc_fit}
   \end{figure*}

As illustrated in Fig.~\ref{fig:massfunc_fit}, the observations suggest dividing the PDMF into four domains. Each domain can be fitted with a power law $m^{-\alpha}$ with the following indices:
\begin{enumerate}
\item $log_{10}(m)\ge$0.2 : $\alpha_{1}$= 2.56  (missing the white dwarfs);
\item -0.7$\le log_{10}(m)\le$-0.22 : $\alpha_{2}$= 0.23;
\item -1.3$\le log_{10}(m)\le$-0.7 : $\alpha_{3}$= -1.22;
\item -1.45$\le log_{10}(m)\le$-1.3 : $\alpha_{4}$= 0.13.
\end{enumerate}

\subsection{The mass function in time: comparison to other clusters}

\begin{figure*}
   \centering
   \includegraphics[width=0.95\textwidth]{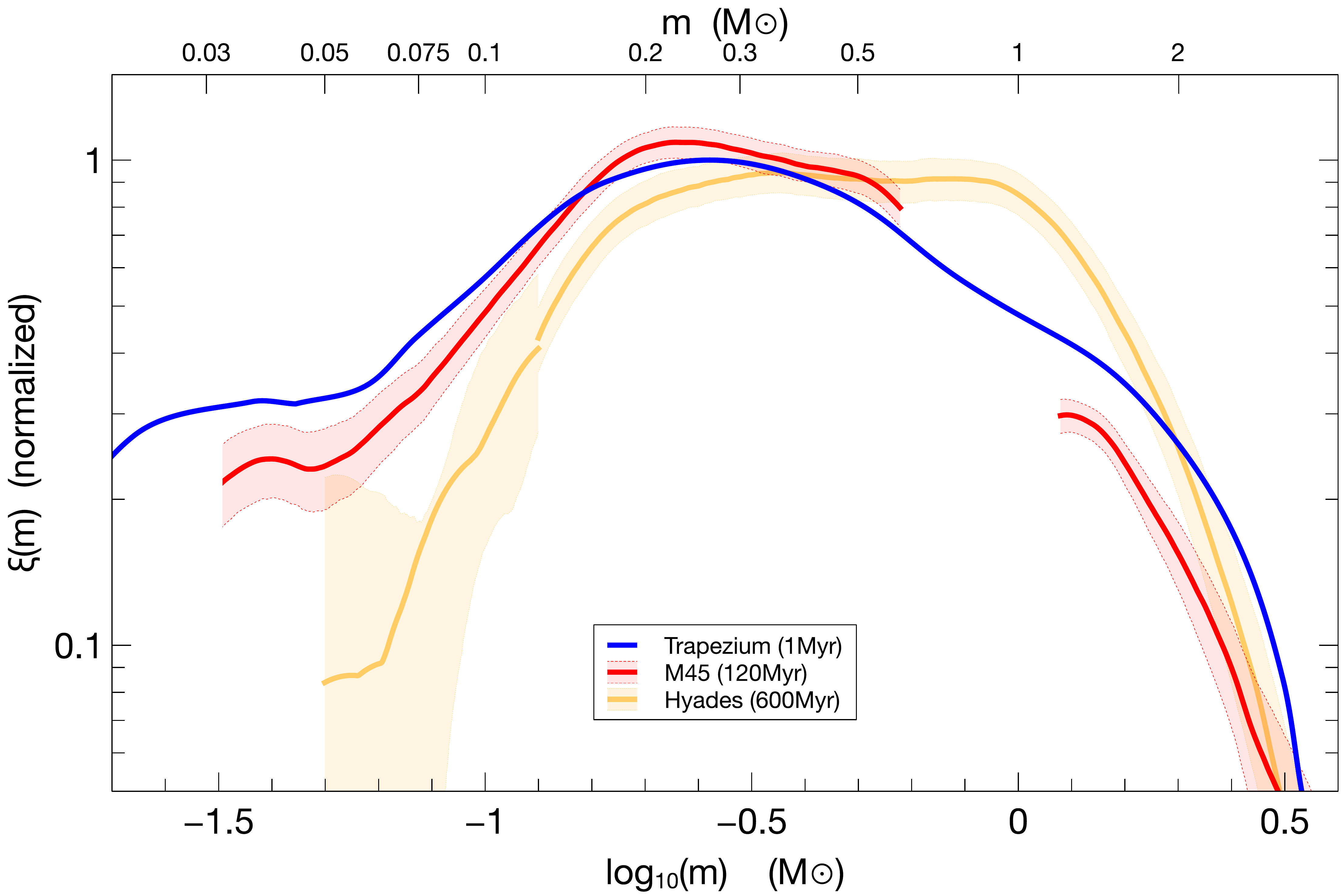}
      \caption{Observed present-day system mass function of the Pleiades within the central 3\degr\, radius (red), compared to the measurements reported for the Trapezium cluster by \citet{2002ApJ...573..366M} and \citet{2000ApJ...540..236H} (blue) and the Hyades (orange) from \citet{2013A&A...559A..43G} (down to 0.15~M\sun) and \citet{2008A&A...481..661B} in the range 0.05$\le$M$\le$0.15~M\sun.}
         \label{fig:massfunc4}
   \end{figure*}

Figure~\ref{fig:massfunc4} compares the results of our study to the PDMF reported in the young $\sim$1~Myr old Trapezium cluster and in the 625~Myr old Hyades cluster.
The Hyades PDMF was computed as follows. The stellar PDMF was computed using the sample of members and masses reported in \citet{2013A&A...559A..43G} and the substellar PDMF using the sample of brown dwarfs and masses reported in \citet{2008A&A...481..661B}. In both cases a KDE with an Epanechnikov kernel with a bandwidth of 0.1 (in $log_{10}(m)$) was used. The two studies covered very different areas. The substellar PDMF obtained from the \citet{2008A&A...481..661B} sample was therefore scaled to match the integrated mass of the \citet{2013A&A...559A..43G} sample over the mass range 0.1--0.15~M$_{\sun}$ where both studies are expected to be complete and overlap. 

The PDMF of the Trapezium cluster was estimated by transforming the $K$-band luminosity function reported by \citet{2002ApJ...573..366M} into a mass function using the BT-Settl models at 1~Myr up to 1~M$_{\sun}$, and \citet{2012MNRAS.427..127B} at 1~Myr models above 1~M$_{\sun}$.

The Trapezium and Hyades clusters were not chosen randomly for the comparison. They are similar to the Pleiades in many aspects. Their total masses and numbers of members are similar, and various studies have derived their mass functions over a similar mass domain, making the comparison meaningful and interesting. Additionally, the Pleiades has been described as a snapshot of the Trapezium's future \citep{2001MNRAS.321..699K} and of the Hyades' past \citep{2008A&A...481..661B}. Figure~\ref{fig:massfunc4} therefore illustrates the evolution of the mass function in time.

In this context and as already known \citep[see, e.g.,][ and references therein]{2012scel.book..115M}, Figure~\ref{fig:massfunc4} shows that the proportion of very low mass stars and brown dwarfs is clearly decreasing over time and that the peak of the mass function is shifting toward higher masses from 0.2~M$_{\sun}$ at the age of the Trapezium and of the Pleiades to 0.5$\sim$1~M$_{\sun}$ by the age of the Hyades. 

The flattening of the Pleiades PDMF below 0.05$\sim$0.06~M$_{\sun}$ is observed in the Trapezium PDMF as well, reinforcing the idea that the overabundance of low mass brown dwarfs is due to additional formation mechanisms -- including  photo-evaporation, disk instabilities, and ejections -- rather than late ($>$1~Myr) dynamical evolution. A similar flattening might be present in the Hyades PDMF around 0.075~M$_{\sun}$, but the very large uncertainties due to the very-small-number statistics in that mass range prevent any firm conclusion and meaningful comparison.  

In any case, and within the large uncertainties of each PDMF at the very low mass end, the level of overabundance around 0.05~M$_{\sun}$ seems to be decreasing with time, since it is rather large at the age of the Trapezium and the Pleiades and possibly shifted toward higher masses (if present) at the age of the Hyades. This would be consistent with the classical picture in which a significant number of very low mass stars and brown dwarfs are ejected before the first million years, and subsequently evaporate as the cluster becomes more and more relaxed.

At the high mass end, the comparison is not straightforward. Both the Hyades and Pleiades samples miss the white dwarfs. Including the white dwarfs would certainly change the comparison. The Trapezium includes several massive stars. By the age of the Pleiades, the most massive of them (which fall beyond Fig.~\ref{fig:massfunc4} coverage) have since long exploded as supernov\ae\,. Some massive stars might also have been ejected from their parent multiple stellar system. In any case, the lack of statistics at the high mass end means that the comparison of the mass function over the corresponding domain is often not statistically significant.

\section{Conclusions}

We have presented a study of the Pleiades cluster's properties. We combined the results of the DANCe survey \citep[][updated with new photometry and improved membership probabilities]{2014A&A...563A..45S} with a reanalysis of the Tycho-2 catalog over the same area. Our analysis allows us to:
\begin{enumerate}
\item identify 2\,010 candidate members in the DANCe survey; 
\item identify 207 candidate members in the Tycho-2 catalog, including 87 new members;
\item provide the largest and most complete census of the cluster. After removing matched entries in the DANCe and Tycho-2 samples (110 sources) and adding Electra and Alcyone (missed in both surveys), the final sample of Pleiades members adds to 2\,109 members, including 812 previously unknown members;
\item estimate contamination levels between 35\% and 61\% in previous surveys from the literature searching for Pleiades brown dwarfs; 
\item provide accurate empirical isochrones in six broad band optical and near-infrared filters ($r$,$i$,$Y$,$J$,$H$,$Ks$);
\item show that theoretical isochrones poorly match the empirical sequence in most color-magnitude diagrams made of optical and near-infrared colors and luminosities;
\item observe that the luminosity function becomes flat beyond $Ks\ga$15.5~mag, corresponding to $\sim$0.04 - 0.05~M$_{\sun}$ and suggesting that different or additional formation mechanisms must be at work at the very low mass end;
\item show that the overall shape of the theoretical models or predictions of the system initial mass function is in reasonable agreement with the observed system's present-day mass function (assuming that the theoretical mass-luminosity relationship is correct). They predict too many brown dwarfs, but the uncertainty on the mass-luminosity relationships prevents us from drawing any firm conclusion about this difference. They do not predict the flattening down to 0.05~M$_{\sun}$
\end{enumerate}

\begin{acknowledgements}
We are grateful to our anonymous referee for a constructive report that helped significantly improve our original manuscript.
We are grateful to I. Thies and P. Kroupa for providing the electronic table of their Pleiades mass function predictions. We are grateful to L. Hillebrand and G. Muench for providing an electronic version of the mass and luminosity functions of the Trapezium cluster presented in their respective articles.

H. Bouy is funded by the the Ram\'on y Cajal fellowship program number RYC-2009-04497. This research has been funded by Spanish grants AYA2012-38897-C02-01 and AYA2010-21161-C02-02. E. Moraux ackowledges funding from the Agence Nationale pour la Recherche program ANR 2010 JCJC 0501 1 ``DESC (Dynamical Evolution of Stellar Clusters)''. J. Bouvier acknowledges funding form the Agence Nationale pour la Recherche program ANR 2011 Blanc SIMI 5-6 020 01 (``Toupies''). This publication makes use of VOSA, developed under the Spanish Virtual Observatory project supported from the Spanish MICINN through grant AyA2008-02156. Based on observations obtained with MegaPrime/MegaCam, a joint project of {\it CFHT} and CEA/DAPNIA, at the Canada-France-Hawaii Telescope (CFHT), which is operated by the National Research Council (NRC) of Canada, the Institut National des Science de l'Univers of the Centre National de la Recherche Scientifique (CNRS) of France, and the University of Hawaii. Based on observations obtained with WIRCam, a joint project of CFHT, Taiwan, Korea, Canada, and France, at the Canada-France-Hawaii Telescope (CFHT), which is operated by the National Research Council (NRC) of Canada, the Institute National des Sciences de l'Univers of the Centre National de la Recherche Scientifique of France, and the University of Hawaii. This paper makes use of data obtained from the Isaac Newton Group Archive, which is maintained as part of the CASU Astronomical Data Centre at the Institute of Astronomy, Cambridge. The data was made publicly available through the Isaac Newton Group's Wide Field Camera Survey Programme. The Isaac Newton Telescope is operated on the island of La Palma by the Isaac Newton Group in the Spanish Observatorio del Roque de los Muchachos of the Instituto de Astrofísica de Canarias. This research used the facilities of the Canadian Astronomy Data Centre operated by the National Research Council of Canada with the support of the Canadian Space Agency. This research draws upon data provided by C. Briceño as distributed by the NOAO Science Archive. NOAO is operated by the Association of Universities for Research in Astronomy (AURA) under cooperative agreement with the National Science Foundation. This publication makes use of data products from the Two Micron All Sky Survey, which is a joint project of the University of Massachusetts and the Infrared Processing and Analysis Center/California Institute of Technology, funded by the National Aeronautics and Space Administration and the National Science Foundation. This work is based in part on data obtained as part of the UKIRT Infrared Deep Sky Survey.  This research has made use of the VizieR and Aladin images and catalog access tools and of the SIMBAD database, operated at the CDS, Strasbourg, France. We are grateful to the Isaac Newton Group for the service observations with LIRIS at the WHT. The William Herschel Telescope and its service program are operated on the island of La Palma by the Isaac Newton Group in the Spanish Observatorio del Roque de los Muchachos of the Instituto de Astrofísica de Canarias. This publication makes use of data products from the Wide-field Infrared Survey Explorer, which is a joint project of the University of California, Los Angeles, and the Jet Propulsion Laboratory/California Institute of Technology, funded by the National Aeronautics and Space Administration. This work is based [in part] on archival data obtained with the Spitzer Space Telescope, which is operated by the Jet Propulsion Laboratory, California Institute of Technology under a contract with NASA. Support for this work was provided by NASA.

\end{acknowledgements}

\bibliographystyle{aa}
\bibliography{mybiblio}

\begin{appendix} 
\section{DANCe-Pleiades Catalogue Data Release 2\label{secondrelease}}
In this new release, the DANCe-Pleiades catalogue presented in \citet{2013A&A...554A.101B} and \citep{2014A&A...563A..45S}  has been improved in the following ways:
\begin{itemize}
\item addition of APASS $gri$ photometry when no measurement was available;
\item addition of $Y$-band photometry for 40 bright candidate members from the list of \citet{2007ApJS..172..663S};
\item recalculation of the membership probability using the new photometry following the method described in \citep{2014A&A...563A..45S}. 
\end{itemize}

Complementing the previous release at the bright end using the APASS  and our own $Y$-band observations was particularly useful for the membership analysis. The selection process described in \citet{2014A&A...563A..45S} made use of the $ri$ and $YJHK$ photometry. The $JHK$ photometry comes from a combination of 2MASS and UKIDSS and covers the entire dynamic range up to $\sim$19~mag. On the other hand, the $riY$ photometry was limited to $\sim$13~mag, and was preventing us from defining a training set of members with complete photometric coverage for the brightest sources. 

To circumvent this problem, we gathered $gri$ photometry from the APASS survey. The APASS DR7 release covers $\sim$97\% of the sky and 100\% of the area encompassed by the DANCe-Pleiades survey in the Johnson $B$ and $V$ and Sloan $gri$ filters. Since no Johnson photometry was available in the DANCe-Pleiades catalog, it was ignored and we only included the $gri$ photometry. The APASS survey is valid from about the 10$^{th}$ to 17$^{th}$ magnitudes. Brighter sources are not as reliable, but we nevertheless chose to report the corresponding photometry in our catalog, keeping this limitation in mind.

No shallow $Y$-band survey was available in the Pleiades at the time of our analysis. Service time was granted to us at the William Herschel Telescope (Proposal SW2013b27, P.I. Bouy) with its {\it LIRIS} near-infrared camera \citep{2002INGN....6...22A}. Shallow $Y$-band images of a sample of 40 bright candidate members of \citet{2007ApJS..172..663S} with available $riJHK$ photometry and randomly selected to cover the luminosity range between 8--14~mag were obtained on 23 December 2013 and 3 January 2014. Sets of five individual exposures of 0.9~s were obtained at five positions around the target.  The ambient condition was good, and the photometric zeropoints were derived by comparison of the faint sources present in the images with their UKIDSS counterparts. The individual raw images were processed using an updated version of \emph{Alambic} \citep{Alambic}, a software suite developed and optimized for the processing of optical and near-infrared imagers, and configured for LIRIS. \emph{Alambic} includes standard processing procedures, such as dark subtraction for each individual readout ports, flat-field correction, bad pixel masking, destripping, and background subtraction.

The analysis presented in \citet{2014A&A...563A..45S} was then repeated including the new photometry. It allowed us to extend the training set, hence the final calculation of membership probabilities,  at the high luminosity end up to $\sim$9~mag. As a result, a total of 97 members originally missed by our analysis were recovered at the bright end. Table~\ref{dance-r2} gives the astrometry, photometry and membership probability for the second release of the DANCe catalogue presented in this study.

\begin{figure}
   \centering
   \includegraphics[width=0.45\textwidth]{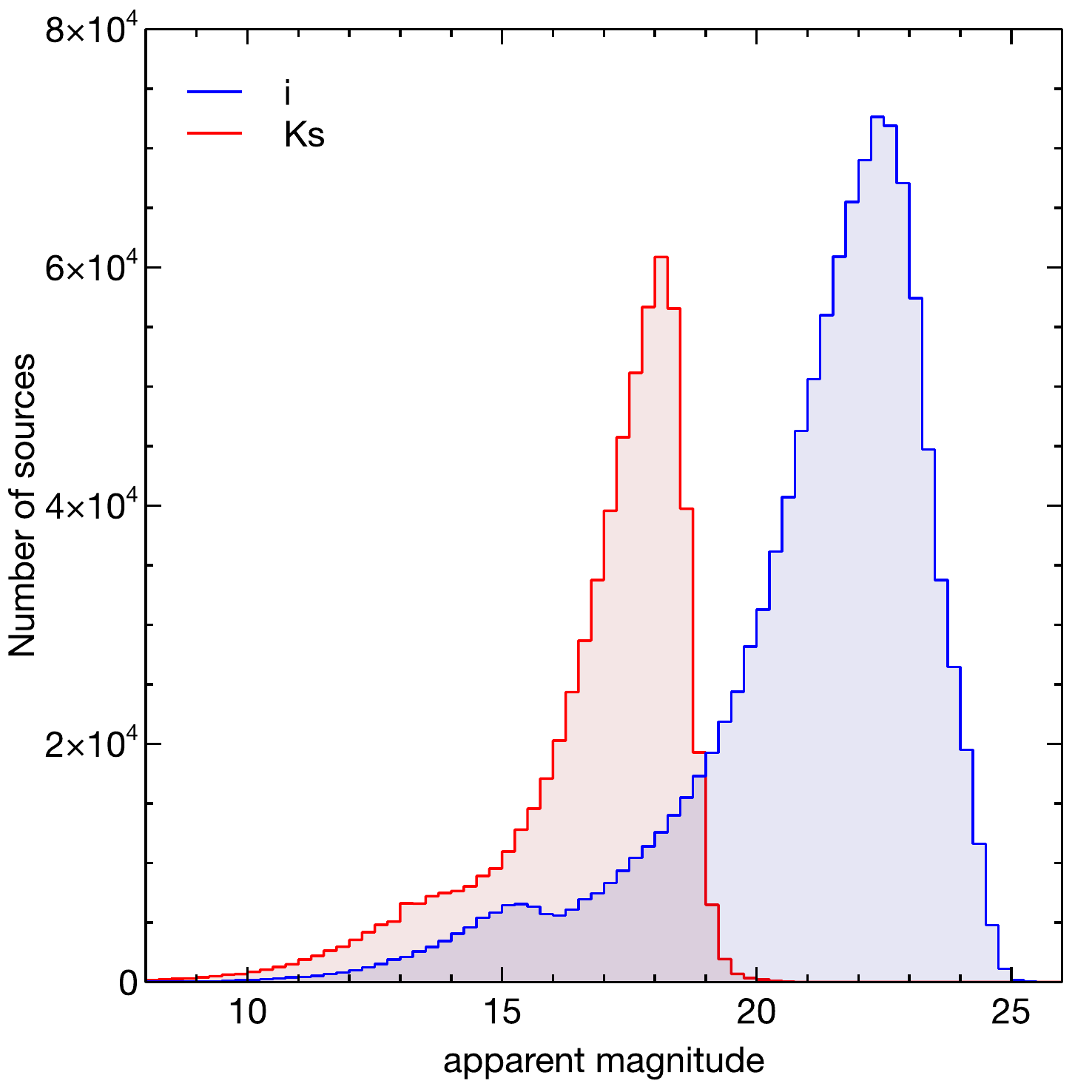}
      \caption{Distribution of apparent magnitudes for all the sources in the survey in the $i$-band (blue) and $Ks$-band (red). }
         \label{fig:sensitivity}
   \end{figure}

Figure~\ref{fig:sensitivity} shows the distribution of apparent magnitudes in the $i$ and $Ks$ bands for the entire sample. These histograms provide a useful estimate of the completeness and sensitivity limits of the survey. In the $i$-band, the survey is sensitive up to 25~mag and complete up to $\sim$23~mag. In the $Ks$-band, the survey is sensitive up to $\sim$20~mag and complete up to $\sim$18~mag. A small excess appears around $i\sim$15~mag and $Ks\sim$14~mag. In the $i$-band, it corresponds to the overlap between APASS and the DANCe photometry, with the APASS survey encompassing a larger area. In the $Ks$-band, it corresponds to the overlap between 2MASS and UKIDSS, which have different spatial resolutions.

\section{Selection of Pleiades members in the Tycho-2 catalog \label{tycho_sel}}
\begin{figure}
   \centering
   \includegraphics[width=0.45\textwidth]{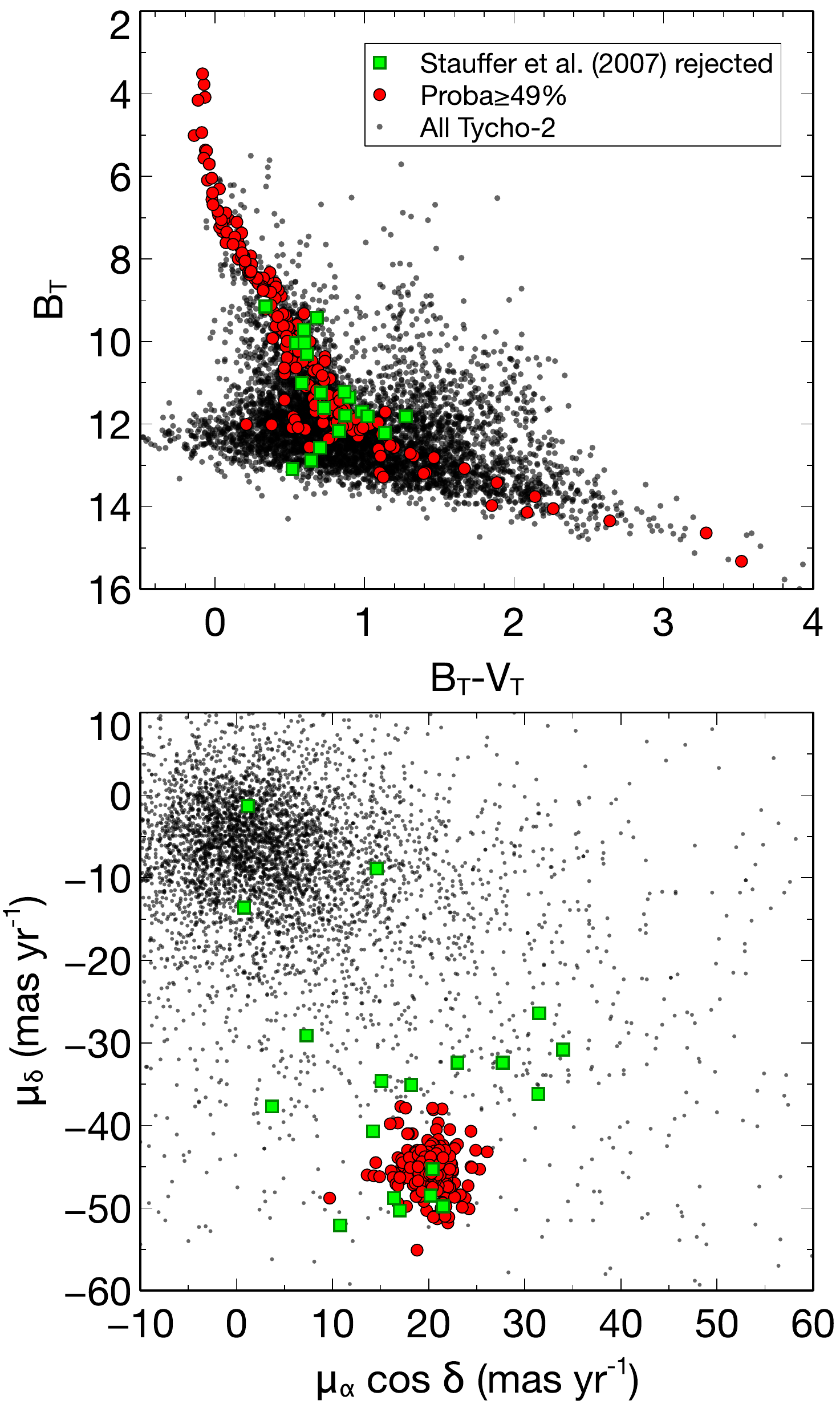}
      \caption{Upper panel: $B_{\rm T}$ vs $B_{\rm T}-V_{\rm T}$ color-magnitude diagram of the Tycho-2 catalog in the area of the DANCe survey (black dots). New candidate members (with membership probability greater than of equal to 48\%) are represented by red dots. Candidate members from \citet{2007ApJS..172..663S} that do not pass the selection criterion are represented by green squares. Lower panel: vector point diagram of the Tycho-2 catalog in the area of the DANCe survey. The color and symbol codes are the same.}
         \label{fig:bv_v_vpd}
   \end{figure}

Although the Tycho-2 catalog has been extensively used to search for Pleiades members, we decided to perform our analysis using the method developed by \citet{2014A&A...563A..45S}. We retrieved the Tycho-2 catalog within the same area as the DANCe survey and complemented it with APASS, 2MASS, CMC-14, and WISE photometry. The best set of variables \citep[in a statistical sense, see ][ for more details]{2014A&A...563A..45S} was found to include the proper motions and the $B_{\rm T}$, $V_{\rm T}$, $r$, J, H, and K magnitudes. The training set was made of 175 candidate members from the list of \citet{2007ApJS..172..663S} with a counterpart in the Tycho-2 catalog. Simulations of the completeness and contamination \citep[see][for more details]{2014A&A...563A..45S} suggest to use a threshold of 48\% of probability to select members in the final catalog. Figure~\ref{fig:bv_v_vpd} shows a $B_{\rm T}$ vs $B_{\rm T}-V_{\rm T}$ and vector point diagrams for the 5\,537 sources included in this study. With this criterion, a total of 207 candidate members are selected. We find that:
\begin{itemize}
\item 83 of the 207 candidate members were not present in \citet{2007ApJS..172..663S} list, and
\item 19 candidate members out of the original 175 of \citet{2007ApJS..172..663S} list with a Tycho-2 counterpart are not selected in our analysis. Figure~\ref{fig:bv_v_vpd} shows that a significant fraction of them are rejected based on inconsistent proper motions, and a few of them are only based on inconsistent colors and luminosities.
\end{itemize}
Table~\ref{dance-tycho2} gives the astrometry, photometry and membership probability for the 5\,537 Tycho-2 sources included in this study.

\end{appendix}
 
\onltab{
\begin{table*}
\caption{Present day system bolometric luminosity function of the Pleiades (normalized) \label{tab:bollumfunc}}
\begin{tabular}{lc}
\hline\hline
$M_{\rm bol}$ ($M\sun$) & Density \\ 
\hline
-3.14 & 0.0018$^{+6.0E-4}_{-0.0015}$ \\
-2.26 & 0.0022$^{+0.001}_{-0.0016}$ \\
-1.38 & 0.0025$^{+0.0015}_{-0.0015}$ \\
-0.5 & 0.0036$^{+0.002}_{-0.0021}$ \\
0.38 & 0.0066$^{+0.0019}_{-0.0032}$ \\
1.26 & 0.011$^{+0.0024}_{-0.0041}$ \\
2.14 & 0.0167$^{+0.002}_{-0.0052}$ \\
3.02 & 0.0215$^{+0.0024}_{-0.0043}$ \\
3.89 & 0.027$^{+0.0017}_{-0.0039}$ \\
4.77 & 0.0277$^{+6.0E-4}_{-0.006}$ \\
7.21 & 0.0831$^{+0.012}_{-0.023}$ \\
8.04 & 0.1088$^{+0.0144}_{-0.027}$ \\
8.86 & 0.1796$^{+0.018}_{-0.048}$ \\
9.69 & 0.2668$^{+0.0195}_{-0.0638}$ \\
10.51 & 0.2447$^{+0.0634}_{-0.0305}$ \\
11.33 & 0.1234$^{+0.0896}_{-0.0226}$ \\
12.16 & 0.0615$^{+0.0573}_{-0.0156}$ \\
12.98 & 0.0388$^{+0.0239}_{-0.0136}$ \\
13.8 & 0.0397$^{+0.0099}_{-0.0137}$ \\
\hline
\end{tabular}
\end{table*}
}

\onltab{
\begin{table*}
\caption{Present day system mass function of the Pleiades (normalized) {\bf assuming  an age of 120~Myr, a distance of 136.2~pc and a mean extinction of $A_{\rm V}=$0.12~mag} \label{tab:massfunc}}
\begin{tabular}{lc}
\hline\hline
$log_{10}(m)$ & $\xi(m)$ \\ 
\hline
-1.55 & 0.18$^{+0.04}_{-0.04}$ \\
-1.41 & 0.24$^{+0.04}_{-0.04}$ \\
-1.26 & 0.25$^{+0.04}_{-0.04}$ \\
-1.11 & 0.34$^{+0.05}_{-0.05}$ \\
-0.96 & 0.54$^{+0.06}_{-0.06}$ \\
-0.82 & 0.85$^{+0.07}_{-0.08}$ \\
-0.67 & 1.08$^{+0.08}_{-0.08}$ \\
-0.52 & 1.05$^{+0.08}_{-0.08}$ \\
-0.38 & 0.96$^{+0.07}_{-0.07}$ \\
-0.23 & 0.81$^{+0.07}_{-0.07}$ \\
0.13 & 0.29$^{+0.02}_{-0.02}$ \\
0.19 & 0.24$^{+0.02}_{-0.02}$ \\
0.25 & 0.19$^{+0.02}_{-0.02}$ \\
0.31 & 0.14$^{+0.02}_{-0.02}$ \\
0.37 & 0.1$^{+0.02}_{-0.02}$ \\
0.43 & 0.07$^{+0.02}_{-0.02}$ \\
0.5 & 0.05$^{+0.02}_{-0.02}$ \\
0.56 & 0.03$^{+0.02}_{-0.01}$ \\
0.62 & 0.02$^{+0.01}_{-0.01}$ \\
0.68 & 0.02$^{+0.01}_{-0.01}$ \\
\hline
\end{tabular}
\end{table*}
}

\onltab{
\begin{table*}
\caption{DANCe catalogue of the Pleiades (second release) \label{dance-r2}}
\begin{tabular}{lrrrrrrrrrrrrrrrrrrrrrrrrrrrrrrrrrrlrrrr}
\hline
DANCe &
Sarro+2014 &
NPOS\_TOTAL &
NPOS\_OK &
RA &
Dec &
err\_RA &
err\_Dec &
PMALPHA\_J2000 &
PMDELTA\_J2000 &
PMALPHAERR\_J2000 &
PMDELTAERR\_J2000 &
CHI2\_ASTROM &
EPOCH &
EPOCH\_MIN &
EPOCH\_MAX &
u &
g &
r &
i &
z &
Y &
J &
H &
K &
err\_u &
err\_g &
err\_r &
err\_i &
err\_z &
err\_Y &
err\_J &
err\_H &
err\_K &
proba &
Model &
Teff &
e\_Teff &
Lbol &
e\_Lbol \\
\hline
  J035008.43+253255.5 & 5403154 & 44 & 44 & 57.5351086587683 & 25.5487606105672 & 3.3705095E-7 & 3.3705095E-7 & 14.023258 & -41.923634 & 0.96409976 & 0.96409976 & 2.686213 & 2010.46655941246 & 1997.88067077344 & 2011.09854136956 & 20.659254 & 18.3969974517822 & 17.1300086975097 & 15.5905838012695 & 14.6606323623657 & 13.7146358489990 & 13.1507968902587 & 12.6089119338989 & 12.3024957275390 & 0.006817403 & 0.00124887598212 & 7.91207654401... & 0.00130840076599 & 0.05017279272644 & 0.05004319919821 & 0.05004236122238 & 0.04010088063365 & 0.06002094278862 & 0.99933670683620 & bt-settl & 3200.0 & 50.0 & 0.013303465 & 0.0021899198\\
  J034623.03+243617.8 & 5237969 & 80 & 80 & 56.5959755649232 & 24.6049563409070 & 2.4088035E-7 & 2.4088035E-7 & 18.390747 & -38.316517 & 0.3549254 & 0.3549254 & 2.8220322 & 2008.87691664123 & 1998.88761615331 & 2011.09856643459 & 21.045883 & 18.7518367767334 & 17.4464607238769 & 15.9762430191040 & 15.0748484039306 & 14.1224889755249 & 13.5631885528564 & 13.0270929336547 & 12.6670987701416 & 0.017129727 & 0.00139840040355 & 8.05286341346... & 6.30847061984... & 0.05026493934408 & 0.05006242909464 & 0.05005712817085 & 0.05002833254323 & 0.06001686902251 & 0.99934060370997 & bt-settl & 3200.0 & 50.0 & 0.009364503 & 0.0015398115\\
  J034524.79+242045.1 & 5157061 & 75 & 75 & 56.3532988755393 & 24.3458727345082 & 2.5063028E-7 & 2.506179E-7 & 15.696801 & -37.347153 & 0.36873412 & 0.3684971 & 2.0417693 & 2008.78593613632 & 1999.96086625598 & 2011.09856643459 & 20.80952 & 18.3822212219238 & 17.0047416687011 & 15.5464019775390 & 14.6229841613769 & 13.6770458221435 & 13.1181240081787 & 12.5389995574951 & 12.1803529357910 & 0.016815327 & 0.00132487167138 & 6.67744374368... & 5.57460705749... & 0.05014828993022 & 0.05003870750806 & 0.05003690879930 & 0.03299999982118 & 0.06001068311813 & 0.99934078779755 & bt-settl & 3200.0 & 50.0 & 0.014233927 & 0.0023434153\\
  J035254.91+243718.1 & 5290218 & 46 & 46 & 58.2288026911685 & 24.6216882239095 & 1.2403241E-6 & 1.2403241E-6 & 15.284061 & -41.780487 & 1.2234616 & 1.2234616 & 1.0174732 & 2008.30965115852 & 1996.94318959616 & 2010.80643256876 &  &  &  & 23.6521511077880 &  &  & 18.8031845092773 & 17.7541694641113 & 16.9255638122558 &  &  &  & 0.03307389467954 &  &  & 0.12105540507466 & 0.08755656132666 & 0.07681207593299 & 0.99934175618144 & bt-settl & 1700.0 & 50.0 & 1.1016685E-4 & 2.118751E-5\\
  J034133.72+250648.5 & 5224174 & 41 & 41 & 55.3904898811303 & 25.1134754192527 & 2.0917902E-7 & 2.091718E-7 & 16.158274 & -39.303555 & 0.16774064 & 0.16774234 & 2.8445265 & 2002.26927966575 & 1998.98260442162 & 2011.09858417114 & 20.188988 & 17.9366340637207 & 16.6679191589355 & 15.3092193603515 & 14.3838093185424 & 13.5017719268798 & 12.9770002365112 & 12.3360004425048 & 12.0241706466674 & 0.00798049 & 0.00121098791714 & 0.00108324608299 & 0.00140790140721 & 0.05010840665127 & 0.05003430420389 & 0.02400000020861 & 0.02999999932944 & 0.06001306137599 & 0.99934264216402 & bt-settl & 3300.0 & 50.0 & 0.01680734 & 0.0027639738\\
  J034043.20+224953.8 & 4991637 & 45 & 45 & 55.1799876390125 & 22.8316197776090 & 1.6703784E-7 & 1.6703784E-7 & 17.586416 & -39.40968 & 0.11161399 & 0.11161399 & 3.8043845 & 2005.09438066264 & 1998.75139876796 & 2011.09859763751 & 21.371082 & 19.0863780975341 & 17.7688064575195 & 16.1558132171630 & 15.2474710845947 & 14.2946472167968 & 13.7302694320678 & 13.1769037246704 & 12.8488881683349 & 0.015813228 & 0.00209541525691 & 0.00140266621019 & 0.00212890468537 & 0.05031574055553 & 0.05007361491257 & 0.05007933448527 & 0.05004164307780 & 0.06004961528822 & 0.99934811558531 & bt-settl & 3200.0 & 50.0 & 0.007806687 & 0.0012867078\\
  J034809.22+235840.5 & 5167474 & 146 & 146 & 57.0384110265741 & 23.9779110883695 & 2.0908E-7 & 2.0907564E-7 & 14.156092 & -38.23998 & 0.27931514 & 0.27940235 & 2.8715425 & 2008.30096769951 & 1997.88062067077 & 2011.09856643459 & 20.8157 & 18.6214160919189 & 17.3031864166259 & 15.8278675079345 & 14.9345238113403 & 14.0237646102905 & 13.4475994110107 & 12.8794664764404 & 12.5840213394165 & 0.01043809 & 0.00101380934938 & 5.68124523852... & 4.64829994598... & 0.05022662707159 & 0.05005666713636 & 0.05005200801710 & 0.04001802783237 & 0.06000922900539 & 0.99935040252490 & bt-settl & 3200.0 & 50.0 & 0.010265237 & 0.0016903129\\
  J034744.44+243655.6 & 5238429 & 108 & 108 & 56.9351582236966 & 24.6154514381828 & 5.5228713E-7 & 5.5228713E-7 & 15.799539 & -37.738228 & 0.553527 & 0.553527 & 0.8266261 & 2007.67725852971 & 1996.94045174538 & 2011.09856643459 & 21.5225 & 19.3189868927001 & 18.0642204284667 & 16.4750423431396 & 15.5751564407348 & 14.5547523498535 & 13.9936122894287 & 13.4472398757934 & 13.1324348449707 & 0.028687846 & 0.00171401922125 & 0.00100880756508 & 5.85996021982... & 0.05046761382940 & 0.05009696765448 & 0.05008685110891 & 0.05004616925324 & 0.05006071298871 & 0.99935100678095 & bt-settl & 3100.0 & 50.0 & 0.0063279322 & 0.0010403749\\
  J034505.00+234606.4 & 5150226 & 83 & 83 & 56.2708149165531 & 23.7684469977015 & 2.404183E-7 & 2.4041105E-7 & 16.7521 & -40.920746 & 0.1825553 & 0.18254629 & 1.8462653 & 2004.91803629153 & 1998.85793319644 & 2011.09857051937 & 21.991125 & 19.6236515045166 & 18.2588272094726 & 16.4075622558593 & 15.4093143844604 & 14.3077430725097 & 13.6867895126342 & 13.1738014221191 & 12.7754371261596 & 0.03332603 & 0.00271794619038 & 0.00142104434780 & 6.99735828675... & 0.05037790416125 & 0.05007282324847 & 0.05006776689482 & 0.05003455646330 & 0.06003795496139 & 0.9993526857756 & bt-settl & 3000.0 & 50.0 & 0.007865492 & 0.0012962425\\
  J034651.44+240616.0 & 5149477 & 94 & 94 & 56.7143228114129 & 24.1044497268851 & 2.8111455E-7 & 2.8110395E-7 & 17.048769 & -38.71549 & 0.37770113 & 0.37780273 & 1.2467431 & 2007.88221013174 & 1998.96939903518 & 2011.08768346371 & 21.451607 & 19.5886249542236 & 18.4285945892334 & 16.7547512054443 & 15.7769195938110 & 14.7773818969726 & 14.1964836120605 & 13.6239814758300 & 13.2719259262084 & 0.058858573 & 0.00455991784110 & 0.00139979319646 & 6.72850874252... & 0.05058305415544 & 0.05011449725943 & 0.05010507012115 & 0.05005644336461 & 0.05006615299983 & 0.99935434995495 & bt-settl & 3000.0 & 50.0 & 0.0054001645 & 8.8400545E-4\\
  J034336.58+231233.9 & 4997214 & 49 & 49 & 55.9024017319216 & 23.2094137491369 & 3.154203E-7 & 3.154203E-7 & 17.193773 & -39.260647 & 0.69039243 & 0.69039243 & 1.4574896 & 2010.30493856949 & 1998.85788336755 & 2011.09859248020 & 20.85899 & 18.6608409881591 & 17.3427162170410 & 15.7213468551635 & 14.6978933715820 & 13.7808752059936 & 13.2029123306274 & 12.6314329528808 & 12.3264167404174 & 0.011439838 & 0.00177377706859 & 0.00115768611431 & 0.00136015319731 & 0.05016818653550 & 0.05004461271535 & 0.05004087520485 & 0.04002433807495 & 0.06000865971152 & 0.99935672366230 & bt-settl & 3100.0 & 50.0 & 0.012587609 & 0.0020747434\\
  J034342.89+255137.0 & 5383859 & 46 & 46 & 55.9287136644351 & 25.8602804825768 & 2.2304917E-7 & 2.2304917E-7 & 16.92722 & -43.731045 & 0.17735845 & 0.17735845 & 2.0717864 & 2004.67728254352 & 2000.73102368240 & 2011.09858009632 & 21.993607 & 19.6493301391601 & 18.3440113067626 & 16.6177940368652 & 15.6810371780395 & 14.6946315765380 & 14.0979595184326 & 13.5251150131225 & 13.2016630172729 & 0.03243444 & 0.00347406999208 & 0.00218389742076 & 0.00420267554000 & 0.05052805133647 & 0.05012333434507 & 0.05011986989238 & 0.05006191294046 & 0.05006355151647 & 0.99935872542277 & bt-settl & 3100.0 & 50.0 & 0.0055912565 & 9.1803446E-4\\
  J034411.28+245234.1 & 5227884 & 61 & 61 & 56.0470181374588 & 24.8761386932101 & 4.078212E-7 & 4.078212E-7 & 15.970753 & -40.609146 & 0.7464518 & 0.7464518 & 1.211094 & 2009.79874727736 & 1999.90059685147 & 2011.09856235157 & 22.41501 & 19.8821926116943 & 18.5349998474121 & 16.8559246063232 & 15.8638050460815 & 14.9035320281982 & 14.2801551818847 & 13.7265634536743 & 13.4341526031494 & 0.037164364 & 0.00356784812174 & 0.00190145452506 & 0.00134311325382 & 0.05064285374854 & 0.05014598945979 & 0.05012027165881 & 0.05008042173714 & 0.05004769026326 & 0.99936484090142 & bt-settl & 3100.0 & 50.0 & 0.004685162 & 7.6937093E-4\\
  J035054.65+242155.9 & 5157473 & 94 & 94 & 57.7276985353782 & 24.3655351189338 & 1.2143116E-7 & 1.2142974E-7 & 14.580158 & -42.642223 & 0.08946699 & 0.089467615 & 5.5639887 & 2001.65920523178 & 1996.94045174538 & 2011.09854543512 & 20.981127 & 18.8235836029052 & 17.5262222290039 & 15.9984235763549 & 15.0641653442382 & 14.1491661071777 & 13.5699691772460 & 13.0483560562133 & 12.7571990585327 & 0.012391975 & 0.00130992277991 & 7.76341650635... & 5.56634622626... & 0.05026596764386 & 0.05006067588715 & 0.05005816758625 & 0.05002890370421 & 0.06003129628358 & 0.99936668130553 & bt-settl & 3200.0 & 50.0 & 0.008975641 & 0.0014773393\\
  J035244.28+235415.1 & 5171639 & 42 & 42 & 58.1845187631600 & 23.9042076976708 & 1.9325248E-7 & 1.9324902E-7 & 14.417653 & -40.78907 & 0.13090026 & 0.13090025 & 1.8627511 & 2001.06007801454 & 1996.94045174538 & 2011.09852467899 & 22.825624 & 20.5362644195556 & 19.1608123779296 & 17.2903270721435 & 16.2281172180175 & 15.1838006973266 & 14.5538606643676 & 14.0002450942993 & 13.6543006896972 & 0.053747084 & 0.00733427563682 & 0.00376676628366 & 0.00133497826755 & 0.05090701445568 & 0.05019609415460 & 0.05017813861640 & 0.05012317235599 & 0.05012952198828 & 0.99937109397059 & bt-settl & 3000.0 & 50.0 & 0.0035148663 & 5.789985E-4\\
  J034227.31+223424.5 & 4935291 & 39 & 39 & 55.6138046074025 & 22.5734844985916 & 5.0468094E-7 & 5.0461955E-7 & 15.373779 & -37.72428 & 0.83277494 & 0.83526933 & 2.0418205 & 2009.55358428560 & 1998.75144558521 & 2011.09859763751 & 19.712389 & 17.8373222351074 & 16.4866371154785 & 14.9796218872070 & 14.1416361236572 & 13.2231388092041 & 12.6346053695678 & 12.1125301742553 & 11.8092000579834 & 0.007099399 & 0.00174275867175 & 0.00123462488409 & 0.00225110701285 & 0.05007722679619 & 0.05002544004827 & 0.04001901383368 & 0.04003469037123 & 0.06001431003962 & 0.99937145782055 & bt-settl & 3200.0 & 50.0 & 0.021454848 & 0.0035285172\\
  J034427.50+241416.9 & 5152272 & 73 & 73 & 56.1145974564091 & 24.2380192856017 & 2.511189E-7 & 2.5110643E-7 & 16.914013 & -41.516357 & 0.37660876 & 0.376349 & 2.8987212 & 2008.85083833953 & 1999.90059822039 & 2011.09856643459 & 21.284285 & 19.1093902587890 & 17.7912960052490 & 16.0734195709228 & 15.1234352493286 & 14.1012153625488 & 13.5026445388793 & 12.9151234054565 & 12.5739829635620 & 0.021866476 & 0.00189185305498 & 0.00103786191903 & 6.14656659308... & 0.05025540926941 & 0.05005748308453 & 0.05005466651219 & 0.04008795156250 & 0.06002340280735 & 0.99937270902609 & bt-settl & 3100.0 & 50.0 & 0.00956645 & 0.001579111\\
  J034024.19+243504.0 & 4134727 & 56 & 56 & 55.1008096693649 & 24.5844530665781 & 1.6823185E-7 & 1.682281E-7 & 18.037254 & -39.45197 & 0.11775634 & 0.11775787 & 5.9336114 & 2004.05119422866 & 1998.98306458015 & 2011.09861005014 & 19.230957 & 16.9243030548095 & 15.6614942550659 & 14.5390892028808 & 13.7534048461914 & 13.0363436508178 & 12.4264077758789 & 11.8299999237060 & 11.5782144165039 & 0.0025266854 & 5.79322513658... & 5.77375991269... & 0.00185224111191 & 0.05004476128327 & 0.03002772880104 & 0.04001619919456 & 0.02999999932944 & 0.06001103775688 & 0.99937607439404 & bt-settl & 3500.0 & 50.0 & 0.028479353 & 0.0046726214\\
  J034741.45+235818.7 & 5154110 & 78 & 78 & 56.9227179675443 & 23.9718482388656 & 2.6196128E-7 & 2.6196128E-7 & 17.636019 & -38.43552 & 0.35440356 & 0.35440356 & 5.6662674 & 2008.59549975691 & 1998.88764380561 & 2011.08768346371 & 15.88407 & 14.0570001602172 & 13.0210000991821 & 12.5249996185302 &  &  & 11.0659999847412 & 10.4610004425048 & 10.3120002746582 & 7.7586615E-4 & 0.04399999976158 & 0.10440306508910 & 0.05099999904632 &  &  & 0.01999999955296 & 0.01700000092387 & 0.01999999955296 & 0.99937769187626 & Kurucz & 4250.0 & 125.0 & 0.13693486 & 0.02232907\\
  J034357.28+241320.4 & 5146806 & 94 & 92 & 55.9886710359267 & 24.2223212907073 & 1.8217466E-7 & 1.8217466E-7 & 16.246382 & -38.73788 & 0.14212586 & 0.14212586 & 4.969371 & 2004.56698303101 & 1998.97185048761 & 2011.09858826799 & 20.731138 & 18.4853610992431 & 17.1717910766601 & 15.6285676956176 & 14.6960699462890 & 13.7751617431640 & 13.1901960372924 & 12.6162609481811 & 12.3090293502807 & 0.015298386 & 0.00126650137826 & 8.21767200250... & 5.35873929038... & 0.05015833526145 & 0.05005015097664 & 0.05004011825319 & 0.04012988004447 & 0.06000854139090 & 0.99937886078493 & bt-settl & 3200.0 & 50.0 & 0.012965756 & 0.002135531\\
  J035041.20+253930.6 & 5399682 & 24 & 24 & 57.6716797131958 & 25.6584928110123 & 2.8959272E-7 & 2.8958004E-7 & 17.924747 & -42.85534 & 0.24511959 & 0.24510232 & 1.8701829 & 2003.27260733801 & 1997.88067871320 & 2011.08765434911 & 22.038166 & 19.9897136688232 & 18.5995960235595 & 16.8346652984619 & 15.8423330688476 & 14.7908477783203 & 14.1669540405273 & 13.6140069961547 & 13.2723722457885 & 0.030622484 & 0.00486943079158 & 0.00243287044577 & 0.00132860534358 & 0.05056392872793 & 0.05013007618400 & 0.05011307340266 & 0.05007217278882 & 0.05008051357648 & 0.99938177404898 & bt-settl & 3000.0 & 50.0 & 0.0051833694 & 8.5069163E-4\\
  J034011.92+255232.3 & 5374365 & 60 & 60 & 55.0496801381824 & 25.8756419920711 & 2.8825687E-7 & 2.8825687E-7 & 17.354225 & -38.844566 & 0.21420881 & 0.21420881 & 2.2720387 & 2006.07614968554 & 1999.85993921971 & 2011.09860191625 & 22.962584 & 20.4168968200683 & 19.0260601043701 & 17.2879409790039 & 16.1922485733032 & 15.2307529449462 & 14.5890855789184 & 14.0577468872070 & 13.7004575729370 & 0.03941486 & 0.00458230171352 & 0.00229789107106 & 0.00629015685990 & 0.05125669270667 & 0.05022071663094 & 0.05023026893539 & 0.05011338157781 & 0.05014678631774 & 0.99938302099970 & bt-settl & 3000.0 & 50.0 & 0.0034575372 & 5.683359E-4\\
  J034412.14+235237.2 & 5146567 & 89 & 89 & 56.0506014764596 & 23.8770056221031 & 2.2743258E-7 & 2.2743258E-7 & 19.524122 & -39.724197 & 0.22859783 & 0.22859783 & 1.9743159 & 2007.43790912095 & 2002.82409308692 & 2011.09858826799 & 20.939253 & 18.7062969207763 & 17.4097633361816 & 15.8236742019653 & 14.9315655136108 & 13.9724111557006 & 13.4079809188842 & 12.8429273986816 & 12.5240256881713 & 0.009106492 & 0.00114330882206 & 7.78598536271... & 6.16964534856... & 0.05022951945185 & 0.05006343206548 & 0.05005171606608 & 0.04004406934272 & 0.06001236793565 & 0.99938323075813 & bt-settl & 3200.0 & 50.0 & 0.010731712 & 0.0017646818\\
  J035558.84+245740.0 & 5294511 & 60 & 60 & 58.9951560925242 & 24.9611100221980 & 1.9141787E-7 & 1.9141787E-7 & 17.867092 & -43.57078 & 0.15322693 & 0.15322693 & 3.442967 & 2004.34233950570 & 1998.78388528405 & 2011.09852060591 & 20.749788 & 18.4233951568603 & 17.1327610015869 & 15.5198564529418 & 14.5564682388305 & 13.5753526687622 & 12.9656047439575 & 12.4358724975585 & 12.1240699386596 & 0.008908334 & 0.00155154685489 & 9.97996306978... & 0.00254847481846 & 0.05011916173434 & 0.05003997305512 & 0.04002285980602 & 0.04008906722662 & 0.06000733246762 & 0.99938476041810 & bt-settl & 3100.0 & 50.0 & 0.015531413 & 0.0025555035\\
\hline
\end{tabular}
\end{table*}
}

\onltab{
\begin{table*}
\caption{DANCe / Tycho-2 catalogue of the Pleiades \label{dance-tycho2}}
\begin{tabular}{lrrrrrrrrrrrrrrrrrrrrrrlrrrrrrrrrrrrr}
\hline
  \multicolumn{1}{c}{DANCe} &
  \multicolumn{1}{c}{RA} &
  \multicolumn{1}{c}{DEC} &
  \multicolumn{1}{c}{pmRA} &
  \multicolumn{1}{c}{pmDE} &
  \multicolumn{1}{c}{e\_pmRA} &
  \multicolumn{1}{c}{e\_pmDE} &
  \multicolumn{1}{c}{BTmag} &
  \multicolumn{1}{c}{e\_BTmag} &
  \multicolumn{1}{c}{VTmag} &
  \multicolumn{1}{c}{e\_VTmag} &
  \multicolumn{1}{c}{r} &
  \multicolumn{1}{c}{err\_r} &
  \multicolumn{1}{c}{g} &
  \multicolumn{1}{c}{err\_g} &
  \multicolumn{1}{c}{i} &
  \multicolumn{1}{c}{err\_i} &
  \multicolumn{1}{c}{J} &
  \multicolumn{1}{c}{e\_J} &
  \multicolumn{1}{c}{H} &
  \multicolumn{1}{c}{e\_H} &
  \multicolumn{1}{c}{K} &
  \multicolumn{1}{c}{e\_K} &
  \multicolumn{1}{c}{Qflg} &
  \multicolumn{1}{c}{W1} &
  \multicolumn{1}{c}{e\_W1} &
  \multicolumn{1}{c}{W2} &
  \multicolumn{1}{c}{e\_W2} &
  \multicolumn{1}{c}{W3} &
  \multicolumn{1}{c}{e\_W3} &
  \multicolumn{1}{c}{W4} &
  \multicolumn{1}{c}{e\_W4} &
  \multicolumn{1}{c}{proba} &
  \multicolumn{1}{c}{Teff} &
  \multicolumn{1}{c}{e\_Teff} &
  \multicolumn{1}{c}{Lbol} &
  \multicolumn{1}{c}{e\_Lbol} \\
\hline
  J033115.95+251519.7 & 52.81646668 & 25.25546943 & 22.2 & -45.3 & 0.8 & 0.8 & 8.44 & 0.016 & 8.165 & 0.013 & 9.179 & 0.07100000232458 & 9.751 & 0.105 & 8.417 & 0.095 & 7.587 & 0.019 & 7.507 & 0.021 & 7.457 & 0.018 & AAA & 7.416 & 0.026 & 7.45 & 0.021 & 7.445 & 0.018 & 7.273 & 0.118 & 0.99863135604317 & 7250.0 & 125.0 & 6.8874235 & 1.1645863\\
  J033128.34+214918.9 & 52.86809418 & 21.82190912 & 23.8 & -48.8 & 0.7 & 0.7 & 9.352 & 0.021 & 8.884 & 0.016 & 9.223 & 0.02899999916553 & 9.791 & 0.034 & 8.769 & 0.004 & 8.071 & 0.023 & 7.89 & 0.017 & 7.824 & 0.023 & AAA & 7.808 & 0.024 & 7.84 & 0.021 & 7.791 & 0.018 & 7.818 & 0.185 & 0.87800811500083 & 6500.0 & 125.0 & 3.9232492 & 0.65190184\\
  J033133.63+261555.8 & 52.89010851 & 26.26548691 & 20.4 & -37.9 & 2.2 & 2.4 & 11.62 & 0.09 & 10.75 & 0.068 & 10.577 & 0.1 & 11.047 & 0.041 & 10.367 & 0.012 & 9.514 & 0.02 & 9.222 & 0.022 & 9.068 & 0.017 & AAA & 9.048 & 0.022 & 9.097 & 0.02 & 9.059 & 0.031 & 8.7 & 0.419 & 0.97343536921912 & 5750.0 & 125.0 & 0.85153997 & 0.15070865\\
  J033305.82+220803.2 & 53.27426485 & 22.13422022 & 21.4 & -49.3 & 1.6 & 1.7 & 12.306 & 0.225 & 11.521 & 0.149 & 11.135 & 0.01099999994039 & 11.775 & 0.017 & 10.934 & 0.045 & 9.897 & 0.018 & 9.522 & 0.017 & 9.425 & 0.02 & AAA & 9.401 & 0.022 & 9.458 & 0.021 & 9.415 & 0.037 & 8.866 & 0.491 & 0.96616861807469 & 5250.0 & 125.0 & 0.5079125 & 0.096009046\\
  J033313.88+230023.1 & 53.30782936 & 23.00641262 & 19.5 & -49.2 & 2.5 & 2.6 & 13.42 & 0.38 & 11.535 & 0.118 & 11.668 & 0.1 & 12.441 & 0.06 & 11.377 & 0.042 & 10.259 & 0.02 & 9.787 & 0.021 & 9.669 & 0.014 & AAA & 9.608 & 0.024 & 9.663 & 0.02 & 9.662 & 0.045 &  &  & 0.95663843262248 & 5000.0 & 125.0 & 0.36470547 & 0.06943537\\
  J033401.81+245251.2 & 53.50754459 & 24.88089936 & 20.8 & -42.1 & 1.6 & 1.6 & 11.704 & 0.079 & 10.932 & 0.072 & 10.691 & 0.01099999994039 & 11.236 & 0.049 & 10.506 & 0.025 & 9.582 & 0.022 & 9.269 & 0.022 & 9.178 & 0.018 & AAA & 9.142 & 0.022 & 9.188 & 0.02 & 9.213 & 0.037 &  &  & 0.97633373292395 & 5500.0 & 125.0 & 0.74389607 & 0.12963206\\
  J033407.31+242040.0 & 53.53045754 & 24.34444364 & 20.8 & -43.5 & 0.9 & 0.9 & 10.119 & 0.032 & 9.645 & 0.03 & 9.431 & 0.1 & 10.309 & 0.053 & 9.329 & 0.028 & 8.553 & 0.034 & 8.312 & 0.021 & 8.274 & 0.018 & AAA & 8.259 & 0.022 & 8.283 & 0.02 & 8.255 & 0.022 & 8.451 & 0.322 & 0.99852674940989 & 6250.0 & 125.0 & 2.2662153 & 0.38790053\\
  J033458.66+233148.5 & 53.74443215 & 23.53014847 & 20.2 & -42.5 & 1.0 & 0.9 & 9.281 & 0.022 & 8.891 & 0.019 &  &  &  &  &  &  & 8.11 & 0.024 & 8.02 & 0.015 & 7.957 & 0.023 & AAA & 7.912 & 0.023 & 7.932 & 0.021 & 7.902 & 0.02 & 7.358 & 0.12 & 0.99587548574356 & 6750.0 & 125.0 & 3.8363912 & 0.6320742\\
  J033531.69+224924.9 & 53.8820554 & 22.82358994 & 21.0 & -44.4 & 1.2 & 1.3 & 10.484 & 0.036 & 9.897 & 0.032 & 9.73 & 0.00200000009499 &  &  & 9.602 & 0.026 & 8.831 & 0.026 & 8.599 & 0.024 & 8.541 & 0.017 & AAA & 8.51 & 0.024 & 8.561 & 0.02 & 8.496 & 0.023 & 8.338 & 0.277 & 0.99895811709423 & 6250.0 & 125.0 & 1.7503414 & 0.29371646\\
  J033608.46+272034.7 & 54.03524149 & 27.3429758 & 22.5 & -48.2 & 1.3 & 1.2 & 10.001 & 0.028 & 9.367 & 0.022 & 9.152 & 0.1 &  &  & 9.128 & 0.046 & 8.188 & 0.018 & 7.916 & 0.017 & 7.88 & 0.029 & AAA & 7.848 & 0.024 & 7.876 & 0.022 & 7.842 & 0.019 & 7.872 & 0.195 & 0.92800976888821 & 6000.0 & 125.0 & 2.970774 & 0.50653213\\
  J033610.76+261552.4 & 54.04483034 & 26.26455153 & 9.7 & -48.8 & 2.1 & 2.1 & 10.942 & 0.056 & 10.339 & 0.047 & 10.063 & 0.1 &  &  & 9.894 & 0.004 & 9.043 & 0.023 & 8.784 & 0.026 & 8.713 & 0.018 & AAA & 8.674 & 0.023 & 8.703 & 0.021 & 8.687 & 0.025 & 8.789 & 0.402 & 0.49792142626451 & 5750.0 & 125.0 & 1.2832775 & 0.21951154\\
\hline
\end{tabular}
\end{table*}
}

\end{document}